\documentclass[12pt,a4paper]{article}
\usepackage{amsmath,amssymb,slashed,cancel}
\usepackage{cite,tabularx}
\usepackage{subcaption}
\usepackage{graphicx,color}
\usepackage[normalem]{ulem}
\usepackage{mathtools}
\usepackage{enumerate}
\usepackage{comment}

\allowdisplaybreaks

\usepackage[height=8.85in,width=6.4in]{geometry}
\renewcommand{\baselinestretch}{1.1}
\newcommand\unit[1]{~\mathrm{#1}}  

\newcommand{\GeV}{~\mathrm{GeV}}
\newcommand{\MeV}{~\mathrm{MeV}}
\newcommand\peryear{\mathrm{/year}}

\newcommand\w[1]{_{\mathrm{#1}}}  

\newcommand\package[2][\relax]{\texttt{#2\ifx#1\relax\relax\relax\else\,\linebreak[0]#1\fi}}

\newcommand\dd{\mathrm{d}}

\DeclareMathOperator{\Order}{\mathcal O}

\numberwithin{equation}{section} 
 
\def\beq#1\eeq{\begin{align}#1\end{align}}

\definecolor{BlueViolet}{rgb}{0.2, 0.00, 0.7}
\definecolor{Blue}{rgb}{0.15, 0.00, 0.9}
\usepackage[colorlinks=true,linkcolor=Blue,citecolor=Blue,urlcolor=BlueViolet]{hyperref}

\begin{document}
\begin{titlepage}
\setcounter{page}{0} 

\begin{center}

\vskip .55in

\begingroup
\centering

\hfill {\rm KEK-TH-2493}
\vskip .55in

{\large\bf
Sub-GeV dark matter search at ILC beam dumps\\
}

\endgroup

\vskip .4in

{
  Kento Asai$^{\rm (a,b)}$,
  Sho Iwamoto$^{\rm (c,d,e)}$,
  Maxim Perelstein$^{\rm (f)}$,\\
  Yasuhito Sakaki$^{\rm (g,h)}$, and
  Daiki Ueda$^{\rm (i,j)}$
}

\vskip 0.25in

  \scalebox{0.9}{%
\begingroup\small
\begin{minipage}[t]{0.94\textwidth}
\centering\renewcommand{\arraystretch}{0.88}
\begin{tabular}{c@{\,}l}
$^{\rm(a)}$
& Institute for Cosmic Ray Research (ICRR), The University of Tokyo, Kashiwa,\\
& Chiba 277--8582, Japan \\[1.5mm]
$^{\rm (b)}$
& Department of Physics, Faculty of Engineering Science, Yokohama National\\
& University, Yokohama 240--8501, Japan \\[1.5mm]
$^{\rm (c)}$
& Institute for Theoretical Physics, ELTE E\"otv\"os Lor\'and University,\\
& Budapest H-1117, Hungary \\[1.5mm]
$^{\rm (d)}$
& Department of Physics, National Sun Yat-sen University, Kaohsiung 804351, Taiwan\\[1.5mm]
$^{\rm (e)}$
& Center for Theoretical and Computational Physics, National Sun Yat-sen University,\\
& Kaohsiung, 804351 Taiwan\\[1.5mm]
$^{\rm (f)}$
& Department of Physics, LEPP, Cornell University, Ithaca, NY 14853, USA \\[1.5mm]
$^{\rm (g)}$
& Radiation Science Center, High Energy Accelerator Research Organization (KEK),\\ &Ibaraki
305--0801, Japan \\[1.5mm]
$^{\rm (h)}$
& The Graduate University for Advanced Studies (SOKENDAI),\\
& Hayama 240--0193, Japan\\[1.5mm]
$^{\rm (i)}$
& Theory Center, High Energy Accelerator Research Organization (KEK),\\ &1-1 Oho, Tsukuba, Ibaraki
305-0801, Japan\\[1.5mm]
$^{\rm (j)}$
& Center for High Energy Physics, Peking University, Beijing 100871, China\\
\end{tabular}
\end{minipage}
\endgroup
}

\end{center}

\vskip .2in

\begin{abstract}
\noindent
Light dark matter particles may be produced in electron and positron beam dumps of the International Linear Collider (ILC). We propose an experimental setup to search for such events, the Beam-Dump eXperiment at the ILC (ILC-BDX). The setup consists of a muon shield placed behind the beam dump, followed by a multi-layer tracker and an electromagnetic calorimeter. The calorimeter can detect electron recoils due to elastic scattering of dark matter particles produced in the dump, while the tracker is sensitive to decays of excited dark-sector states into the dark matter particle. We study the production, decay and scattering of sub-GeV dark matter particles in this setup in several models with a dark photon mediator. Taking into account beam-related backgrounds due to neutrinos produced in the beam dump as well as the cosmic-ray background, we evaluate the sensitivity reach of the ILC-BDX experiment. We find that the ILC-BDX will be able to probe interesting regions of the model parameter space and, in many cases, reach well below the relic target.       

\end{abstract}
\end{titlepage}

\setcounter{page}{1}
\renewcommand{\thefootnote}{\#\arabic{footnote}}
\setcounter{footnote}{0}

\begingroup
\renewcommand{\baselinestretch}{1} 
\setlength{\parskip}{2pt}          
\hrule
\tableofcontents
\vskip .2in
\hrule
\vskip .4in
\endgroup
\section{Introduction}
\label{sec:introduction}

The International Linear Collider (ILC) has been proposed as the next energy-frontier facility in particle physics.
The physics program of the ILC is well established~\cite{Baer:2013cma,Fujii:2017vwa,ILCInternationalDevelopmentTeam:2022izu}: it includes measurements of the Higgs boson couplings with unprecedented precision, searches for new physics in rare Higgs decays and other channels, and precise top-mass determination, among other topics.
Most studies of the physics potential of the ILC to date focus on experiments with the detector situated at the main interaction point, where the electron and positron beams collide head-on, yielding the maximum collision energy.

Recently, it was suggested that, in parallel with the experiments at the main interaction point, the ILC can pursue a complementary physics program using its beam dumps~\cite{Kanemura:2015cxa,Sakaki:2020mqb,Asai:2021ehn,Asai:2021xtg,Moroi:2022qwz,Nojiri:2022xqn,Giffin:2022rei}.%
\footnote{%
  In particular, the ILC provides a good opportunity for a positron fixed-target experiment, similar to the PADME experiment~\cite{PADME:2023vvr} but at a much higher energy.
}
Particle collisions within the beam dumps occur in fixed-target kinematics, with typical effective energies of $\Order(10)\GeV$ or below.
At the same time, since every beam-electron and beam-positron interacts with the dump, the beam-dump experiments can accumulate enormous integrated luminosity.
This makes the beam dump the ideal place to search for relatively light new particles with very small couplings to the Standard Model (SM).

The new physics targeted by the beam-dump experimental program is very well motivated theoretically~\cite{Alexander:2016aln,Battaglieri:2017aum}.
Previously studied examples include visibly-decaying dark photons, axion-like particles, vector bosons associated with gauged lepton flavor symmetries, and heavy neutral leptons.
In all these cases, it was demonstrated that beam-dump experiments at the ILC can probe model parameters well beyond the reach of the currently available experiments and are competitive with proposed future dedicated facilities.

In this paper, we propose a new search for light dark matter using the ILC beam dumps.
It is well known that a stable particle with mass in the MeV--GeV range, interacting with the SM via a dark photon mediator, is an attractive dark matter (DM) candidate.
With reasonable model parameters, its thermal relic density matches the observed DM abundance and it is consistent with all astrophysical and experimental constraints.
The DM particles can be pair-produced at the ILC beam dump via virtual dark-photon exchanges or in decays of dark photons.
An electromagnetic calorimeter placed $\sim 100$~m downstream of the beam dump, behind a lead shield is designed to search for the DM by detecting elastic scattering of the DM particles on the electrons in the detector material.
In addition, in many models such as inelastic DM~\cite{Tucker-Smith:2001myb}, the DM particle is the lowest-lying state of a multiplet with small mass splittings. A long-lived excited state in the same multiplet as the DM, produced at the beam dump, may propagate through the lead shield and decay into the DM and charged SM particles. The charged particles can then be detected by a tracker, providing a ``visible decay" signature of the DM.
We will estimate the experimental reach in both electron-recoil and visible-decay signal.

\begin{figure}
\centering
\includegraphics[width=0.95\textwidth]{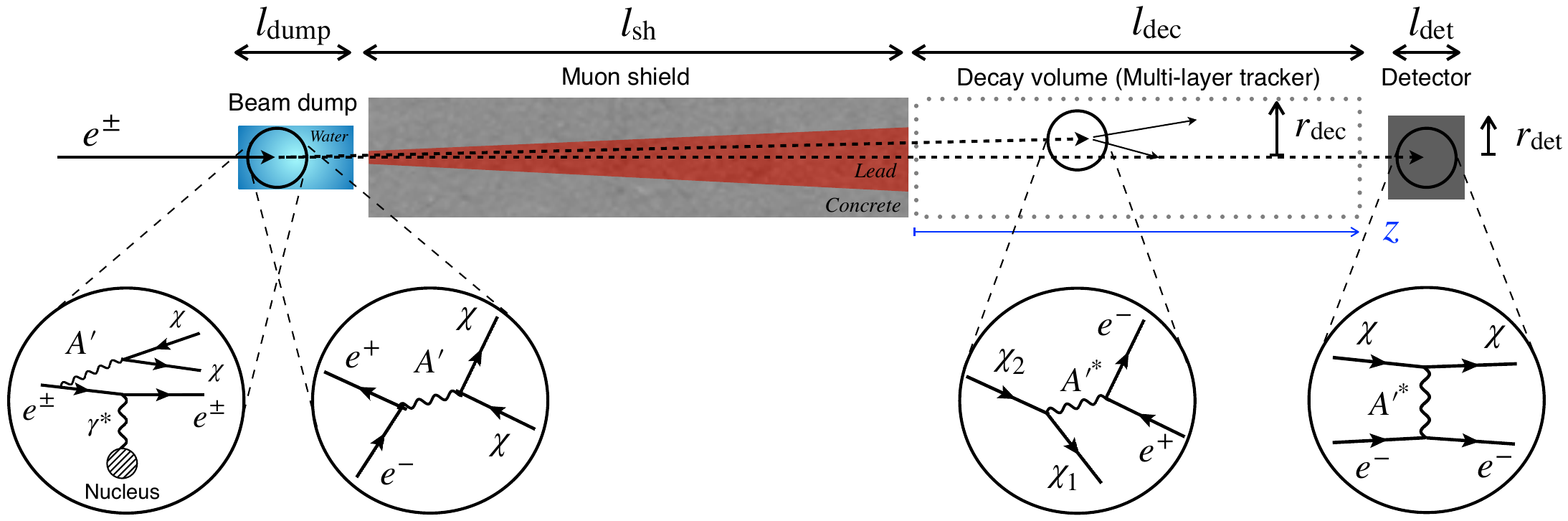}
\caption{
The ILC-BDX experimental setup consists of the main beam dump, a muon shield, a decay volume, and a detector.
A multi-layer tracker is placed in the decay volume to measure the charged tracks.
The DM particles may be produced at electron and positron beam dumps via pair-annihilation of positron on an atomic electron, or via bremsstrahlung.
The resulting DM particles can scatter off electrons in the detector and yield observable electron-recoil events. 
In models where the DM particle is part of a multiplet, an excited DM state may be produced. It can then decay into a lighter DM state and SM particles in the decay volume, producing a visible signature in the tracker. 
}
\label{fig:exp}
\end{figure}

The proposed experimental setup is illustrated in Fig.~\ref{fig:exp}.
The concept of this experiment is the same as the Beam-Dump eXperiment (BDX) proposed at JLab~\cite{BDX:2016akw}, and we, therefore, call our proposal ``ILC-BDX.''
The advantage of the ILC, compared to the original BDX proposal, is the higher beam energy as well as availability of both electron and positron %
beams.%
\footnote{%
  An additional advantage is the availability of polarized beams at the ILC.
  We will not explore consequences of polarization in this paper, but note that it can be particularly useful in characterizing the new physics if a non-SM signal is observed.
}
We also note that the experimental setup used here is essentially identical to that proposed previously for searches for visible dark photons and axion-like particles. (Detailed design of the detector may be different to optimize the sensitivity for various searches, but this is outside the scope of this study.)
The ILC-BDX can be pursued in parallel to the previously discussed searches, adding to the rich physics program at the ILC beam dumps. It is important to emphasize that this program can be pursued in parallel with the exploration of the Higgs and electroweak physics at the main IP, and can broaden the capabilities of the ILC to search for physics beyond the Standard Model (BSM) at a modest additional cost.  

The rest of the paper is organized as follows.
In Section~\ref{sec:setup}, we introduce the setup of our proposed experiment for sub-GeV DM search at the ILC beam dumps, and list formulae used to calculate the number of expected signal events.
In Section~\ref{sec:BG}, we discuss the two types of background (BG) events that occur in this setup, i.e., beam-induced and cosmic-ray backgrounds, and estimate the number of events expected from each BG.
In Section~\ref{sec:Exmaples}, we introduce five DM models used as benchmarks, i.e., pseudo-Dirac DM with small and large mass-splitting, scalar elastic and inelastic DM, and Majorana DM. We then present our estimates of the ILC-BDX sensitivity reach for each model. The main findings of our analysis are summarized in Section~\ref{sec:Summary}.
The Appendix contains the formulae for cross sections of the relevant dark-photon production processes (Appendix~\ref{app:prodcross}) and the DM-electron elastic scattering (Appendix~\ref{app:recoilcross}), as well as some details of our estimates of neutrino-induced BG rates (Appendix~\ref{app:neucross}).

\section{Beam dump experiment}
\label{sec:setup}
%
We adopt the same experimental setup as that of Ref.~\cite{Nojiri:2022xqn}, which is illustrated in Fig.~\ref{fig:exp}.
For both electron and positron beams, the ILC main beam dumps are planned as absorbers consisting of water cylinders along the beam axes with the length of $l_{\rm dump}=11~{\rm m}$~\cite{Satyamurthy:2012zz}.
The proposed setup consists of a muon shield with the length of $l_{\rm sh}=70~{\rm m}$ made of lead, a cylindrical decay volume with $l_{\rm dec}=50~{\rm m}$ and radius of $r_{\rm dec}=3~{\rm m}$ with a multi-layer tracker installed, and a cylindrical detector with the radius of $r_{\rm det}=2~{\rm m}$ and the length of $l_{\rm det}=0.64~{\rm m}$, behind either of the beam dumps.
The multi-layer tracker is designed to detect the visible-decay signal, while the cylindrical detector is assumed to be an electromagnetic (EM) calorimeter made of CsI(Tl) scintillating crystals ($n_{e^-}^{\rm (det)}=1.1\times 10^{24}~{\rm cm}^{-3}$) designed to detect recoil electrons caused by incoming boosted DM particles (electron-recoil signals).

We focus on the 250 GeV ILC (ILC-250) with the beam energy of $E_{\rm beam}=125~{\rm GeV}$ in the lab frame, and the number of incident electrons and positrons into the beam dump of $N_{e^{\pm}}=4\times 10^{21}/{\rm year}$~\cite{Behnke:2013xla,Baer:2013cma,Adolphsen:2013jya,Adolphsen:2013kya,Behnke:2013lya}. 
As studied comprehensively in Refs.~\cite{Sakaki:2020mqb,Asai:2021ehn,Asai:2022zxw,Nojiri:2022xqn}, this setup was found very sensitive to long-lived BSM particles decaying into visible SM particles thanks to its thick shield.
We will see that, with the calorimeter as a recoil-electron detector, it is also sensitive to DM particle production.

We study two types of signal events associated with different production mechanisms of DM particles.
If the DM particles are produced in the water beam dump by a BSM interaction (such as dark photons), they are highly boosted and, passing through the muon shield, scatter off electrons in the calorimeter.
Such electron recoils, detected by the calorimeter, are a typical signal of DM production.
On the other hand, the DM particles may also be produced in in-flight decays of heavier DM-sector particles if, for example, the DM is the lowest-lying state of a multiplet.
The heavier particle, which we denote by $\chi_2$, can pass through the muon shield and then eventually decay into a DM particle $\chi_1$ and SM particles.
If the decay happens in the decay volume, charged tracks may be observed in the tracker.
This visible-decay signature is used as an additional signal of DM production.

In this work, we focus on five DM models with the dark photon mediator to provide a benchmark study. They are pseudo-Dirac-fermion DM with small or large mass-splitting, scalar elastic DM, scalar inelastic DM, and Majorana-fermion DM.
The details of each model are discussed in Section~\ref{sec:Exmaples}.
Among the models, the pseudo-Dirac DM with small mass-splitting, scalar DM, and Majorana DM can be observed as electron-recoil events.
Meanwhile, the pseudo-Dirac DM with large mass-splitting can produce visible-decay events in addition to electron-recoil signature; specifically, due to the pseudo-Dirac nature, a produced dark photon $A'$ decays mainly into $\chi_2\bar\chi_1$ (or $\chi_1\bar\chi_2$) and the heavier DM-sector particle $\chi_2$ may decay visibly in the decay volume.

The number of signal events is schematically given by
\begin{align}
    N_{\rm signal}=N_{e^{\pm}}\times l_{i} \times n_j\times \sigma_{ij\to A'\to\text{DM}}^{}\times {\rm Acc},
\end{align}
where we consider a particle $i$ in the shower interacting with a particle $j$ in the material of the beam dump to produce DM particles through an on-shell dark photon $A'$ as the mediator.
The track length $l_{e^{\pm}}$ of a shower electron and positron is provided in Ref.~\cite{Asai:2021ehn}\footnote{%
  We estimated properties of electromagnetic showers, namely $l_{e^\pm}$ and $\theta_e$ in Eq.~\eqref{eq:theta_e}, by Monte Carlo simulations with {\tt EGS5}~\cite{Hirayama:2005zm} code embedded in {\tt PHITS~3.23}~\cite{Sato:2018}. The results were further validated through a simulation with {\tt Geant4}~\cite{Agostinelli:2002hh}.
}, $n_j$ is the number density of $j$, and ${\rm Acc}$ denotes the detector acceptance discussed below.
More specifically\footnote{In Ref.~\cite{Nojiri:2022xqn}, this calculation scheme is referred to as the coarse-grained integration method.},
\begin{align}
\label{eq:nsig-ann}
    N^{\rm pair}_{\rm signal}&=N_{e^{\pm}} \int \dd E_{e^+} \frac{\dd l_{e^+}}{\dd E_{e^+}}\cdot n_{e^-}\cdot \sigma (e^+ e^- \to A')\cdot{\rm Br}(A'\to \chi \bar{\chi}) \cdot {\rm Acc},
    \\
\label{eq:nsig-brem}
    N^{\rm brems}_{\rm signal}&=N_{e^{\pm}}\sum_{i=e^-,e^+}\int \dd E_i \frac{\dd l_i}{\dd E_i}\cdot n_{\rm N}^{}\int \dd E_{A'} \int^{\pi}_0 \dd\theta_{A'} \frac{\dd^2 \sigma (i {\rm N}\to i A' {\rm N})}{\dd E_{A'} \dd \theta_{A'}}\cdot{\rm Br}(A'\to \chi \bar{\chi})\cdot {\rm Acc}
\end{align}
for pair-annihilation $e^+e^-\to A'$ and bremsstrahlung $e^\pm\mathrm{N}\to e^\pm A'\mathrm{N}$ with a target nucleus $\mathrm{N}$, respectively.
The cross sections $\sigma$ on the right-hand side are provided in Appendix~\ref{app:prodcross}. Here $\theta_{A'}$ is the emission angle of $A'$ with respect to the direction of the $e^{\pm}$ beam in the lab frame.
In all the models we consider, it is assumed that dark photons decay exclusively into the DM-sector particles, i.e., ${\rm Br}(A'\to \chi \bar{\chi})= 1$, with $\chi$ being a DM sector particle ($\chi_1$ and/or $\chi_2$).
This is justified since decays to SM final states are suppressed by a small mixing parameter $\epsilon^2$.

Let us estimate the acceptance for each type of signal.
For a visible-decay event to be observed, the visible decay of $\chi_2$ must occur in the decay volume.
Noting that $A'$ mainly decays into a $\chi_1$-$\chi_2$ pair in the models we consider, we approximate the acceptance by
\begin{equation}
    {\rm Acc}({\rm decay})=
    \int_0^{r_{\rm dec}/(l_{\rm dump}+l_{\rm sh})}\dd\theta_{\chi} \int_0^{l_{\rm dec}} \dd z \frac{\dd P_{\rm ang}}{\dd\theta_{\chi}}\cdot\frac{\dd P_{\rm dec}}{\dd z}\cdot \Theta (r_{\rm dec}-r_{\perp}^{\rm dec}).\label{eq:Acdec}
\end{equation}
Here, we assume that the decay $A'\to\chi_2\bar\chi_1$ (or $\chi_1\bar\chi_2$) is immediate and $\chi_2$ exclusively decays into electrons because of the small mass difference $\Delta\equiv m_{\chi_2}-m_{\chi_1} < 2 m_{\mu}$.
The lab frame distribution of the $\chi_2$ emission angle $\theta_\chi$ is approximated by\footnote{The angular distribution in the CM-frame is given by $\dd P_{\rm ang}/\dd\cos\theta_{\chi}^{\rm CM}=1/2$ when the polarization of the dark photon is averaged.}
\begin{align}
\label{eq:ang-distribution}
    \frac{\dd P_{\rm ang}}{\dd\theta_{\chi}}=\sin\theta_{\chi}\cdot\frac{1}{2}\left(\frac{m_{A'}^{}}{E_{A'}-p_{A'}^{}\cos\theta_{\chi}}\right)^2,
\end{align}
and $\dd P_{\rm dec}/\dd z$ denotes the probability of $\chi_2$ to decay at the position $z$ (the horizontal axis in Fig.~\ref{fig:exp}), i.e.,
\begin{align}
    \frac{\dd P_{\rm dec}}{\dd z}=\frac{1}{l^{(\rm lab)}_{\chi_2}}{\rm exp}\left(-\frac{l_{\rm dump}+l_{\rm sh}+z}{l^{(\rm lab)}_{\chi_2}} \right),
\qquad
    l^{(\rm lab)}_{\chi_2}=\frac{p_{\chi_2}}{m_{\chi_2}}\frac{1}{\Gamma_{\chi_2}}
\end{align}
with $l^{(\rm lab)}_{\chi_2}$ denoting the lab frame decay length of $\chi_2$.
The lab-frame momentum of $\chi_2$ is approximated by $p_{\chi_2}\approx (E_{A'}+p_{A'}^{}\cos\theta_{\chi})/2$, where the mass-splitting is neglected. The momentum of $A'$ is obtained by $p_{A'}^{}=\sqrt{E_{A'}^2-m_{A'}^2}$, and $\Gamma_{\chi_2}$ ($m_{\chi_2}$) is the total decay width (the mass) of $\chi_2$.
The Heaviside function $\Theta$ in Eq.~\eqref{eq:Acdec} governs the radial requirement on the decay position of $\chi_2$, i.e., the radial deviation
\begin{equation}
\label{eq:radial-dev}
     r_{\perp}^{\rm dec}\approx\sqrt{\theta_e^2 +\theta_{A'}^2+\theta_\chi^2 }\cdot (l_{\rm dump}+l_{\rm sh}+z)
\end{equation}
must be smaller than the radius $r_{\rm dec}$ of the multi-layer tracker.
In Eq.~\eqref{eq:radial-dev}, 
$\theta_e$ is the angle of the beam-oriented $e^{\pm}$ with respect to the beam axis,
$\theta_{A'}$ is the production angle of $A'$ and is equal to 0 for pair-annihilation,
and $\theta_\chi$ is the emission angle of $\chi_2$ at the decay of $A'$.
We estimate $\theta_e$ by Monte Carlo simulations \cite{Asai:2021ehn} and use the mean value
\begin{equation}
    \theta_e= 16~{\rm mrad}\cdot {\rm GeV}/E_{e^{\pm}}.
\label{eq:theta_e}
\end{equation}

Similarly, we estimate the acceptance for electron-recoil signal as
\begin{align}
     {\rm Acc}({\rm recoil}) &= \int_0^{r_{\rm det}/(l_{\rm dump}+l_{\rm sh}+l_{\rm dec})}\dd\theta_{\chi} \frac{\dd P_{\rm ang}}{\dd\theta_{\chi}}\cdot \Theta (r_{\rm det}-r_{\perp}^{\rm rec})\cdot P_{\rm recoil},
\end{align}
where the radial deviation is approximately given by
\begin{equation}
\label{eq:radial-dev-recoil}
    r_{\perp}^{\rm rec}=\sqrt{\theta_e^2 +\theta_{A'}^2 +\theta_\chi^2 }\cdot (l_{\rm dump}+l_{\rm sh}+l_{\rm dec}), 
\end{equation}
and $P_{\rm recoil}$ is the probability of electron recoil
\begin{equation}
 P_{\rm recoil}=n_{e^-}^{(\rm det)}l_{\rm det}\int_{E_{e}^-}^{E_e^+}\dd E_R\frac{\dd\sigma_{\rm recoil}}{\dd E_R}\cdot \Theta (E_R -E_{\rm min})
\end{equation}
given by the electron number density $n^{\rm (det)}_{e^-}$ of the detector, the length $l_{\rm det}$ of detector, kinematically allowed maximum (minimum) recoil energy $E_e^+$ ($E_e^-$) given by Eq.~\eqref{eq:minrecoil}, and the effective recoil cross section approximated by
\begin{equation}
 \frac{\dd \sigma_{\rm recoil}}{\dd E_R}\approx
\frac{\dd\sigma (\chi_1 e^- \to \chi_2 e^-)}{\dd E_{R}}+
\frac{\dd\sigma (\chi_2 e^- \to \chi_1 e^-)}{\dd E_{R}}
\exp\left(-\frac{l_{\rm dump}+l_{\rm sh}+l_{\rm dec}}{l_{\chi_2}^{\rm (lab)}}\right)
\end{equation}
for models with $\chi_2$ and
\begin{equation}
 \frac{\dd\sigma_{\rm recoil}}{\dd E_R}\approx2\times\frac{\dd\sigma (\chi e^- \to \chi e^-)}{\dd E_{R}}
\end{equation}
for the other models.
The analytical formulae for the differential cross section of the DM-electron scattering are shown in Appendix~\ref{app:recoilcross}.
Here, the electron recoil energy $E_R$ is required to be larger than the threshold $E_{{\rm min}}= 1~{\rm GeV}$ to reduce BG events.

\begin{figure}
\centering
\includegraphics[width=0.49\textwidth]{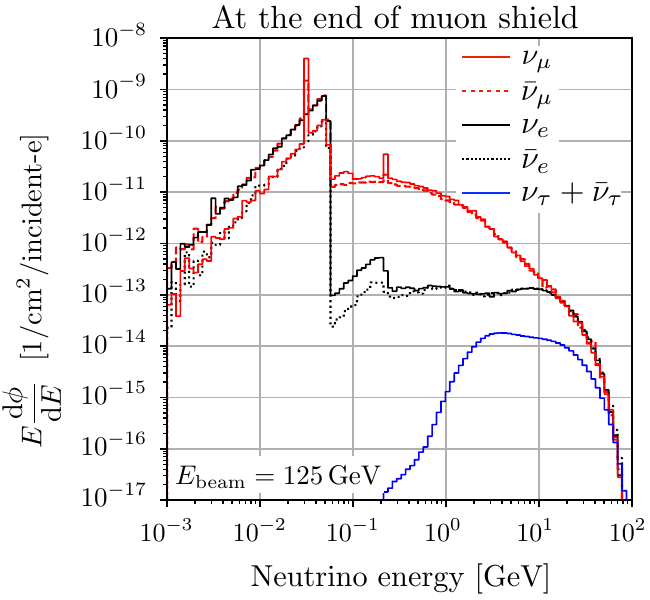}
\includegraphics[width=0.49\textwidth]{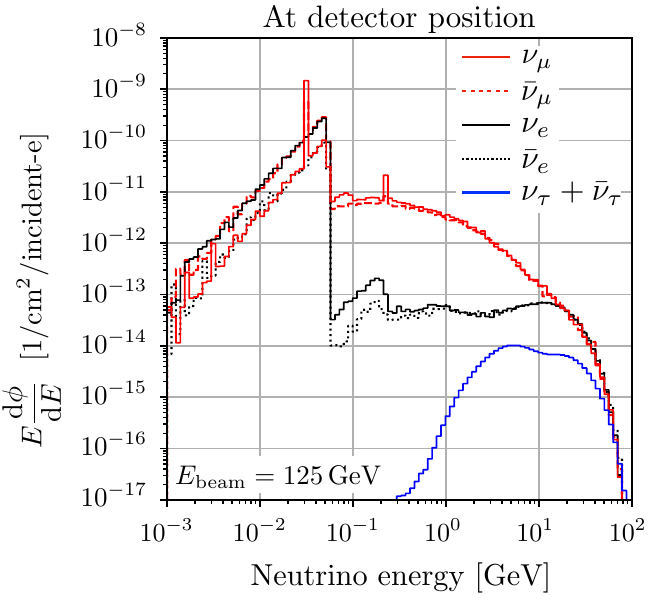}
\caption{
  The beam-induced neutrino fluxes at the end of the muon shield (left) and at the EM calorimeter location (right) behind the electron beam dump.
}
\label{fig:nu}
\end{figure}

\section{Expected backgrounds}
\label{sec:BG}
Background (BG) events in this experiment are classified into two types: beam-induced and beam-unrelated.
The beam-induced BG events arise from the SM particles produced by the injected beam at the main beam dump, while the beam-unrelated ones are mainly due to cosmic-ray muons.

\subsection{Beam-induced background}

The beam injected into the main beam dumps produces SM particles.
In addition to light particles such as pions, neutrons, and muons, because of the initial high-energy beams, tau-leptons and heavy mesons ($D$, $B$, and $B_c$) are also produced by the shower photon hitting nuclei~\cite{Nojiri:2022xqn}\footnote{Heavy meson production through the electromagnetic-shower photons is taken into account, which is
the dominant channel of heavy meson production in our setup.}.
Both light and heavy particle decays can produce neutrinos, which pass through the muon shield and reach the detectors to generate the beam-induced BG events for both the electron-recoil and visible-decay signals.

The neutrino fluxes are calculated with Monte Carlo simulation.
We use PHITS 3.25~\cite{Sato:2018} for production and transport of SM particles other than heavy mesons.
For heavy meson production, the differential production cross sections obtained by PYTHIA 8.3~\cite{Bierlich:2022pfr} are implemented into PHITS; see Ref.~\cite{Nojiri:2022xqn} for details.
In Fig.~\ref{fig:nu}, we show the neutrino fluxes per electron injection at the end of the muon shield in the left panel and at the detector behind the decay volume in the right panel\footnote{%
  At the positron beam dump, the positron annihilation yields additional neutrino fluxes, but differences are negligible above our threshold $E_{\rm min}=1$ GeV.
}.
Tau-type neutrino fluxes are negligible, compared to electron- and muon-types, because the beam energy is not large enough to produce a considerable number of source particles such as tau-leptons and $D_s$.
Due to the stopping power of pions, many pions at rest are produced in the beam dump so that the neutrino flux increases in the low-energy regions.
The electron-type neutrinos in a high-energy regime mainly arise from the decays of $D_s$.

A source of irreducible BG events for the electron-recoil searches are neutrino-electron scattering,\footnote{
  The processes $\nu_\mu e^-\to\mu\nu_e$ and $\nu_\tau e^-\to\tau\nu_e$ do not contribute because the neutrino flux above the threshold energy, $E_{\nu}\geq (m^2_{\mu}-m^2_e)/2m_e\simeq 10.8~{\rm GeV}$ and $E_{\nu}\geq (m^2_{\tau}-m^2_e)/2m_e\simeq 3~{\rm TeV}$, is negligible.
}
    \begin{equation}
        \nu e^- \to \nu e^-,~~~\bar{\nu} e^- \to \bar{\nu} e^-,
    \end{equation}
where beam-originated neutrinos scatter on the atomic electrons in the detector material.
    The recoil electrons produce electromagnetic showers in the calorimeter, which are difficult to distinguish from the DM-electron recoil events.
    From the neutrino flux in the right panel of Fig.~\ref{fig:nu}, the number of neutrino-originated electron-recoil events is estimated to be around $1\peryear$ by imposing $E_{\rm min}=$ 1 GeV; see Appendix~\ref{app:neucross} for details.
Neutrino-nucleon scattering is another source of beam-induced BG for the electron-recoil search.
We list the possible processes below.
Note that, in contrast to the neutrino-electron scattering, these processes provide recoil nucleons in the calorimeter, making this component of the BG potentially reducible depending on the design of the calorimeter.
\begin{itemize}
    \item Quasi-elastic scattering ---
    Neutrinos produced at the beam dump or in the muon shield may interact with material in the calorimeter through the following quasi-elastic scattering processes:
    \begin{align}
    &{\rm CC}:~~~\nu_{\ell} n\to \ell^- p,~~~
    \bar{\nu}_{\ell} p\to \ell^+ n,\notag 
    \\
    &{\rm NC}:~~~\nu p\to \nu p,~~\bar{\nu} p\to \bar{\nu} p,~~\nu n\to \nu n,~~ \bar{\nu} n\to \bar{\nu} n,\notag
    \end{align}
    where CC (NC) denotes charged (neutral) current interaction.
    Among these, the processes $\nu_{e}n\to e^- p$ and $\bar{\nu}_{e}p\to e^+ n$ will produce an electromagnetic shower in the calorimeter and thus potentially be misidentified as an electron-recoil signal.
    The number of such events is conservatively estimated as $2\times 10^2\peryear$ using the simulated neutrino flux (Fig.~\ref{fig:nu} right); see Appendix~\ref{app:neucross} for details. 
    Other processes are without electromagnetic showers, and thus we expect them to be distinguishable from the signal events.

    \item Neutral pion production ---
    In the detector, muon-type neutrinos may produce resonant neutral pions ($\pi^0$) decaying into photons:
    \begin{align}
    {\rm CC}:~~&\nu_{\mu} n\to \mu^- p \pi^0,~~\bar{\nu}_{\mu} p\to \mu^+ n \pi^0,\notag
    \\
    {\rm NC}:~~&\nu_{\mu} p\to \nu_{\mu} p \pi^0,~~
    \bar{\nu}_{\mu} p\to \bar{\nu}_{\mu} p \pi^0,~~\nu_{\mu} n\to \nu_{\mu} n \pi^0,~~\bar{\nu}_{\mu} n\to \bar{\nu}_{\mu} n \pi^0.\notag
    \end{align}
    The electromagnetic showers from the photons, which is accompanied by a recoil nucleon, may be misidentified as electron recoils.
    As detailed in Appendix~\ref{app:neucross}, the number of the single-$\pi^0$ production is conservatively estimated as $2\times 10^3\peryear$ from the flux (Fig.~\ref{fig:nu} right),
    but this BG may be reducible depending on the calorimeter design by, for example, using the information on the radius of the electromagnetic shower.
\end{itemize}
Since these BG events are due to misidentification of electron recoils, we postpone quantitative studies to the detector-design stage.
Instead, in our plots displaying the ILC-BDX sensitivity of the electron-recoil searches, we show curves corresponding to 10, 100, and 1000 signal events in a 10-year ILC run.
These choices illustrate a range of plausible scenarios for the degree to which beam-induced background from neutrino-nucleon scattering can be controlled in practice.

Lastly, we consider the potential beam-induced BG events for the visible-decay signal.
The main source will be the following processes involving strange mesons:
    \begin{align}
       {\rm CC}:~~&
       \nu_{\mu}n \to \mu^- K^0 p,~~\nu_{\mu} n\to \mu^- K^0 \Sigma^+,\notag
       \\
       {\rm NC}:~~&
       \nu_{\mu}n \to \nu_{\mu} K^0 \Lambda^0,~~
       \nu_{\mu}p\to \nu_{\mu} K^0 \Sigma^+,~~\nu_{\mu}n \to \nu_{\mu}K^0 \Sigma^0.\notag
    \end{align}
    Neutral kaons produced at the end of the muon shield and the surrounding walls of the decay volume can decay in the decay volume to yield charged tracks mimicking the visible-decay signal.
    In Ref.~\cite{SHiP:2015vad} of the SHiP experiment, this class of neutrino-induced BG events is evaluated using Geant4~\cite{Allison:2006ve}.
    We infer the number of charged tracks from neutral kaon decays in the decay volume by comparing the number of neutrinos between the ILC beam dump and the SHiP experiment.
    In the SHiP experiment with $2\times 10^{20}$ protons on target, $7\times 10^{17}$ neutrinos with momentum between 2 GeV and 100 GeV are expected, which results in $\sim 10^4$ pairs of charged tracks from neutral kaon decays;
    $99.4$\% of them are rejected by using the topology of oppositely-charged two tracks thanks to the fact that the BG-originated tracks generally do not point to the beam-target interaction points~\cite{SHiP:2015vad}, and around $60$ events are expected to remain as the BG\footnote{The SHiP experiment has a veto system to reduce the BG further. We do not discuss it because the reduction strongly depends on the detector setup.}.
    At the ILC-BDX, the number of neutrinos with momentum between 2 GeV and 100 GeV is estimated as $ 1.4\times 10^{17}$ per 10-year run (cf.\ Fig.~\ref{fig:nu} left), and the ILC-BDX expects 12 BG events with pairs of charged tracks per 10-year run.
    Consequently, in the following analysis of ILC-BDX visible-decay searches,
    we expect 12 BG events and show the expected exclusion limit sensitivity with 95\% confidence-level (C.L.), which corresponds to 7.4 signal events in a 10-year run.

\subsection{Cosmic-ray background}

\begin{figure}
\centering
\includegraphics[width=0.5\textwidth]{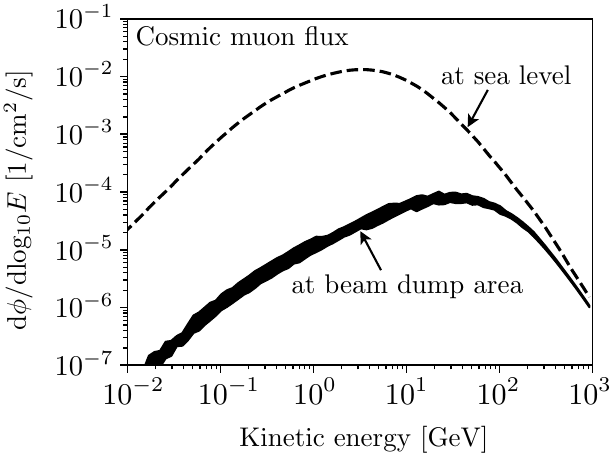}
\caption{
The cosmic-muon fluxes at sea level (dashed line) and the beam dump area (black band). The uncertainly in the latter flux is due to the indeterminate density of subsurface materials in the Kitakami Mountains at the experimental site. The subsurface materials are assumed to be soil and granite, with an average density of 2.0 to 2.4~g/cm$^3$.
}
\label{fig:muon_flux}
\end{figure}

\begin{figure}
\centering
\includegraphics[width=0.7\textwidth]{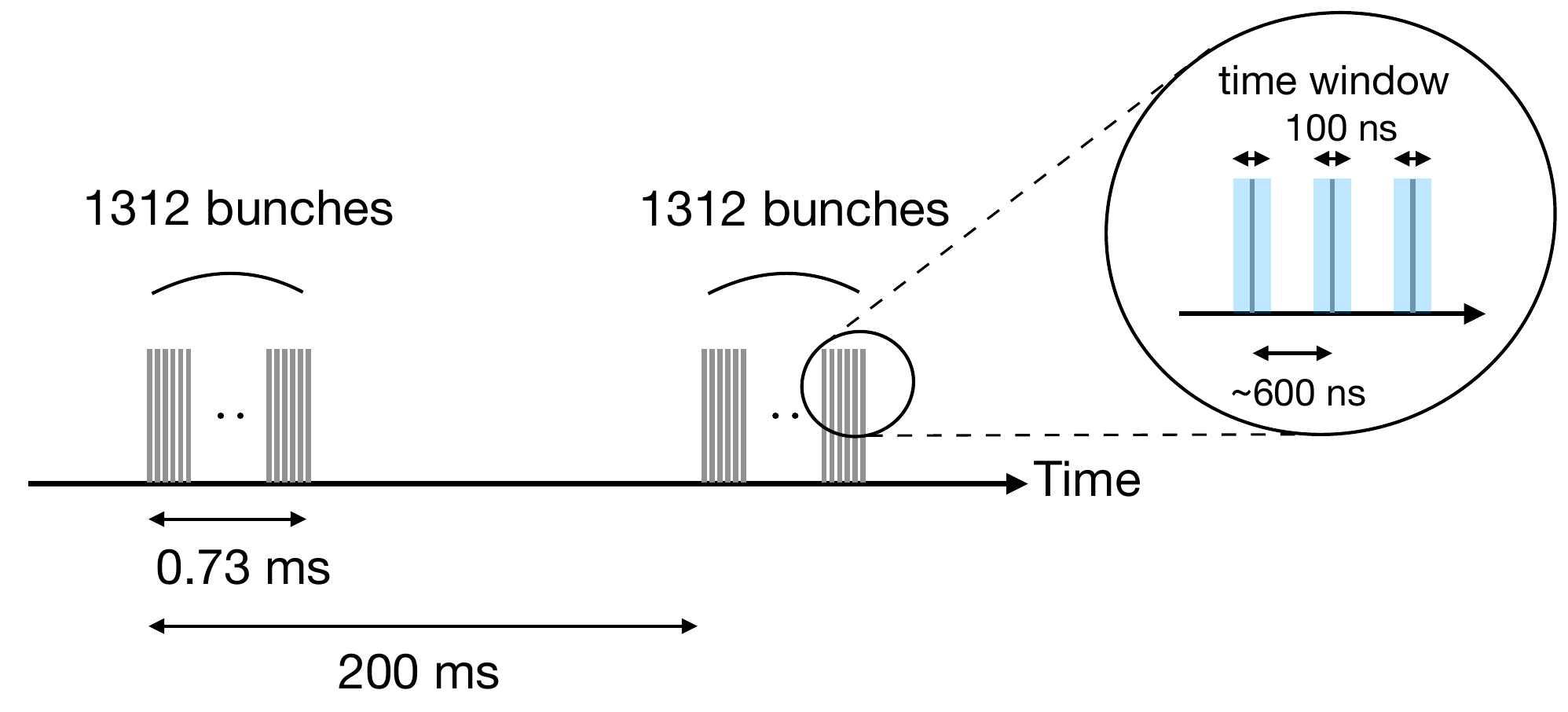}
\caption{
A schematic picture of the ILC beam~\cite{Adolphsen:2013jya,Adolphsen:2013kya}.
The number of bunches per pulse is 1312, the beam pulse length is 0.73 ms, the pulse repetition rate is $1/200~{\rm ms}^{-1}=5\unit{Hz}$, and the bunch spacing is $\sim$600 ns.
The number of the beam-unrelated BG events can be reduced by imposing 100 ns time window per bunch. 
}
\label{fig:beam}
\end{figure}

The beam-unrelated BG is mainly due to cosmic rays.
Figure~\ref{fig:muon_flux} shows the fluxes of cosmic muons at sea level and the beam dump area evaluated by {\tt EXPACS}~\cite{sato2006analytical, sato2008development, sato2015analytical} and {\tt PHITS}~\cite{Sato:2018}\footnote{Muon-induced neutrons avoid the muon veto and may become a BG event. Evaluation of this effect is left for future work.}.
The kinetic energy loss of the cosmic muon from the ground surface to a depth of $\sim$120 meters below the ground surface is estimated as $\rho\times {\langle \dd E/\dd x\rangle}\times 120~{\rm m}\sim 50~{\rm GeV}$ with the mass density of the ground $\rho\sim 2.2~{\rm g/cm^3}$ and the stopping power of muon ${\langle \dd E/\dd x\rangle}\sim 2~{\rm MeV\cdot cm^2/g}$.
Due to this large energy loss, the muon flux at the beam dump area is 200 times smaller than that on the ground.
Then, the number of the cosmic-muon BG events for 10 years is estimated as
\begin{align}
    N^{\rm BG}_{\rm cos}\sim 
    \mathcal{O}(10) \cdot \epsilon_{\rm veto}\label{eq:NBGcos}.
\end{align}
This estimate arises from the following factors:
\begin{align*}
    \mathcal{O}(10) \sim 
    &~ 10~{\rm year}               &\text{(operation time)}\\
    &~\times 10^{-4}~\text{muon}/{\rm cm}^2/{\rm s}  &\text{(cosmic muon flux at beam dump area)}\\
    &~\times 100^2~{\rm cm}^2             &\text{(detector area from top view)}\\
    &~\times 10^{-4}/\text{muon}                  &\text{(hit rate per cosmic muon)}\\
    &~\times 1312\times 5~{\rm bunch}/{\rm s}     &\text{(bunch number per second)}\\
    &~\times 100~{\rm ns}/{\rm bunch}.    &\text{(time window per bunch)}
\end{align*}
The hit rate is the probability that a cosmic-ray muon on an iron block of size $10~{\rm cm} \times 10~{\rm cm} \times 10~{\rm cm}$ will cause an energy deposition above the threshold value of 1~GeV, which is evaluated in the Monte Carlo simulation. 
Detectors with smaller cell sizes are sufficiently realistic, and the above estimation of the hit rate is conservative. The time structure of the ILC bunches~\cite{Adolphsen:2013jya,Adolphsen:2013kya} is represented in Fig.~\ref{fig:beam}.
The factor $\epsilon_{\rm veto}$ is the reduction factor by the cosmic-muon veto, which is typically much smaller than 10\%.
The deep underground location of the detector and coincidence time window significantly reduce the beam-unrelated BG.
Consequently, we will neglect the cosmic-muon BG in the rest of this study.

\section{Examples of detectable DM models}
\label{sec:Exmaples}

We evaluate the sensitivity of the ILC electron and positron beam dump experiments to DM particles using the formulae in Section~\ref{sec:setup}.
Taking into account the beam-induced BG events coming from the SM neutrinos, we show the prospects of the ILC-BDX for electron-recoil and visible-decay searches.
For the electron-recoil searches, we illustrate parameter spaces in which more than 10, $10^2$, and $10^3$ signal events are expected at a 10-year ILC-BDX run.
For the visible-decay searches, regions with more than 7.4 signal events in a 10-year run are shown, which corresponds to a 95\% C.L. exclusion.
Throughout this work, we assume that the beam-unrelated BG events are negligible, as indicated by the estimates in Section~\ref{sec:BG}.

As benchmark models, we consider a class of DM models in which the DM field is charged under a new ``dark'' gauge symmetry, U(1)$_D$.
At low energy, where U(1)$_D$ is spontaneously broken and the dark photon $A'$ acquires a mass $m_{A'}^{}$, the relevant terms of the Lagrangian are given by
\begin{align}
\label{eq:Lag_gauge}
    \mathcal{L}&\supset-\frac{1}{4}F_{\mu\nu}{F}^{\mu\nu}-\frac{1}{4}F'_{\mu\nu}{F'}^{\mu\nu} +\frac{1}{2}m_{A'}^2 A'_{\mu} {A'}^{\mu}-\frac{\epsilon}{2}F'_{\mu\nu}F^{\mu\nu}-g_D^{} A'_{\mu} J^{\mu}_{\chi}-e A_{\mu} J^{\mu}_{\rm EM}~,
\end{align}
where $F_{\mu\nu}$ ($F'_{\mu\nu}$) is the photon (dark photon) field strength, $g_D^{}$ is the U(1)$_D$ gauge coupling, $J^{\mu}_{\chi}$ ($J^{\mu}_{\rm EM}$) is the DM (electromagnetic-matter) current, and
$\epsilon$ parametrizes the kinetic mixing between the photon and the dark photon.
By the redefinition $A_{\mu}\to A_{\mu}-\epsilon A'_{\mu}$, the gauge kinetic terms become canonical, and the interaction terms read
\begin{equation}
 \mathcal L\w{int}=-g_D^{} A'_{\mu} J^{\mu}_{\chi}+\epsilon e A'_\mu J_{\rm EM}^\mu-e A_{\mu} J^{\mu}_{\rm EM}~.
\end{equation}
We consider five models for the nature of the DM: pseudo-Dirac-fermion DM with small or large mass-splitting, scalar elastic DM, scalar inelastic DM, and Majorana-fermion DM.
The DM current $J_\chi^\mu$ for each of these models is listed below, and the recoil profiles for $\chi e \to \chi e$ scattering in the lab frame are summarized in Appendix~\ref{app:recoilcross}.

The DM models used in this study have four or five free parameters: the dark photon mass $m_{A'}^{}$, the DM mass $m_{\chi_{1}}$, the dark fine structure constant $\alpha_D^{} \equiv g_D^2/(4\pi)$, the kinetic mixing parameter $\epsilon$, and the mass difference $\Delta=m_{\chi_2}-m_{\chi_1}$ of DMs if $\chi_2$ is present.
The DM particles remain in chemical equilibrium with the SM plasma in the early universe before freezing out. The DM relic density is determined by the cross section of the pair-annihilation process $\text{DM}+\text{DM}\leftrightarrow\text{SM}+\text{SM}$.
For $m_{\chi}\ll m_{A'}^{}$, the annihilation cross section for these models can be parametrized as $\sigma v\propto y/m_{\chi}^2$ with $y\equiv \epsilon^2 \alpha_D^{} (m_{\chi}/m_{A'}^{})^4$. The cross section for which the relic density matches the observed value defines the {\it ``relic target''} in the $(m_\chi, y)$ space. Following the common practice, we will use this two-dimensional parameter space to illustrate the sensitivity of the ILC beam dump searches. 
Note however that the signal event rates at the ILC-BDX do not depend exclusively on these two parameters and thus further assumptions are necessary to represent the reach on the $(m_\chi,y)$ plane. 
These assumptions will be specified in the captions of each of our reach plots.

\subsection{Pseudo-Dirac DM}
Pseudo-Dirac inelastic DM\footnote{%
  If the DM $\chi$ is a Dirac fermion, the DM current is given by $J^{\mu}_{\chi}\propto\bar{\chi}\gamma^{\mu}\chi$, and it annihilates to SM in $s$-wave two-to-two processes.
  Constraints from the cosmic microwave background (CMB) power injection have ruled out $s$-wave two-to-two annihilating DM lighter than $\mathcal O(10)\GeV$ as the thermal DM candidate~\cite{Leane:2018kjk}. This motivates pseudo-Dirac DM as the simplest viable model of sub-GeV fermionic DM.
  As another possibility, 
  an inelastic dark matter model consisting of two Dirac fermions has also been 
  proposed~\cite{Filimonova:2022pkj}.
}%
~is described by a pair of two-component Weyl fermions $(\eta,\xi)$ that have opposite unit charge under U(1)$_D$. Both a U(1)$_D$-conserving Dirac mass $m_D$ and a U(1)$_D$-breaking Majorana mass $m_M$ are present in the low-energy theory, since the U(1)$_D$ symmetry is spontaneously broken. Namely, the Lagrangian has the mass terms
\begin{align}
    -\mathcal{L}\supset m_D^{} \eta \xi +\frac{1}{2}m_M^{} (\eta^2+\xi^2)+{\rm H.c.}\label{eq:masspseudo}
\end{align}
Here and below, we assume $m_D^{}\gg m_M^{} > 0$.
The mass eigenstates are then given by
\begin{align}
    \chi_1= \frac{i}{\sqrt{2}}(\eta-\xi),~~~\chi_2= \frac{1}{\sqrt{2}}(\eta+\xi)
\end{align}
with masses $m_{\chi_{1,2}}= m_D^{}\mp m_M^{}$, and the DM current becomes off-diagonal:
\begin{align}
    J^{\mu}_{\chi}=i\bar{\chi}_2\gamma^{\mu}\chi_1+{\rm H.c.}
\end{align}

The ILC-BDX search strategy depends on the mass difference $\Delta \equiv m_{\chi_2}-m_{\chi_1} = 2m_M^{}$.
For $\Delta>2 m_e$ (large mass-splitting), the heavier DM particle $\chi_2$ can decay into $\chi_1e^-e^+$ with the partial decay width~\cite{Izaguirre:2017bqb,Giudice:2017zke}
\begin{align}
    \Gamma(\chi_2\to \chi_1 e^- {e}^+)\simeq \frac{4\epsilon^2 \alpha \alpha_D^{} \Delta^5}{15\pi m^4_{A'}}\,,
\end{align}
which results in a visible-decay signal.
Otherwise, if $\Delta<2m_e$ (small mass-splitting), $\chi_2$ does not decay inside the apparatus because of the small width of the main decay channel~$\chi_2\to\chi_1+3\gamma$.
This region of parameter space is accessible only by searches for the electron-recoil signal.

\begin{figure}[t]
\centering
  \begin{subfigure}{0.49\textwidth}%
  \captionsetup{margin={25pt,0pt}}%
  \includegraphics[width=\textwidth]{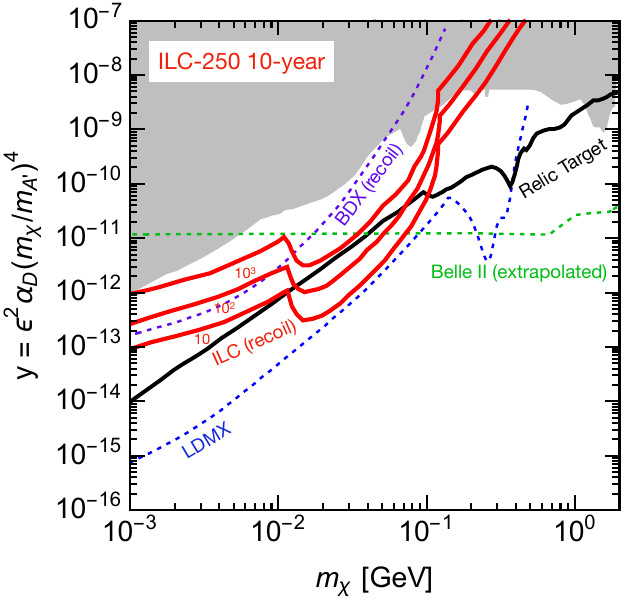}%
  \subcaption{electron beam dump}
  \end{subfigure}%
  \begin{subfigure}{0.49\textwidth}%
  \captionsetup{margin={25pt,0pt}}%
  \includegraphics[width=\textwidth]{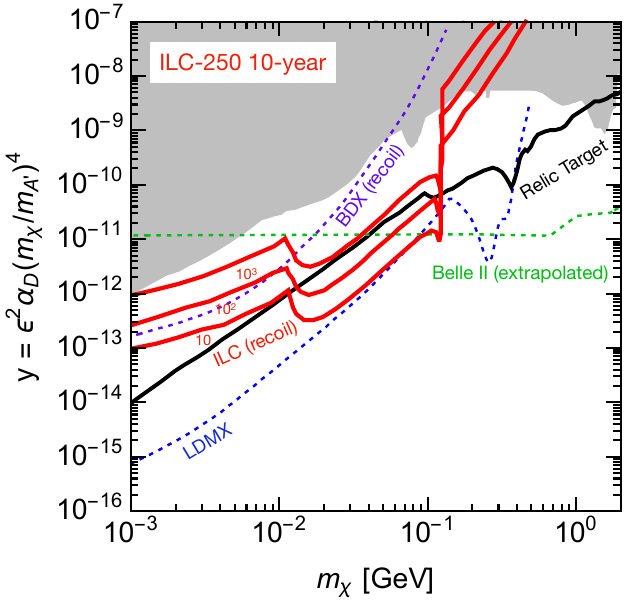}%
  \subcaption{positron beam dump}
  \end{subfigure}%
  \caption{
  Projected sensitivity reach of a 10-year ILC-250 run in the pseudo-Dirac DM model with small mass-splitting, $\Delta \ll\min(m_e,m_{\chi_1})$.
It is assumed that $\alpha_D^{}=0.5$ and $m_{A'}^{}=3m_{\chi}$.
On each panel, the three red solid lines show the sensitivity of the ILC-BDX recoil-electron search, corresponding to $10, 10^2$, and $10^3$ signal events.
The black solid line shows DM relic targets~\cite{Marsicano:2018glj}.
The shaded grey region is excluded by the past experiments; see Section~\ref{sec:cons}. 
The dashed lines show the sensitivity of the BDX recoil-electron search (purple)~\cite{BDX:2016akw}, LDMX (blue)~\cite{Akesson:2022vza}, and Belle~II experiment (green)~\cite{Battaglieri:2017aum,Berlin:2018bsc}.
}
\label{fig:Dirac}
\end{figure}

\begin{figure}[t]
\centering
  \begin{subfigure}{0.49\textwidth}%
  \captionsetup{margin={25pt,0pt}}%
  \includegraphics[width=\textwidth]{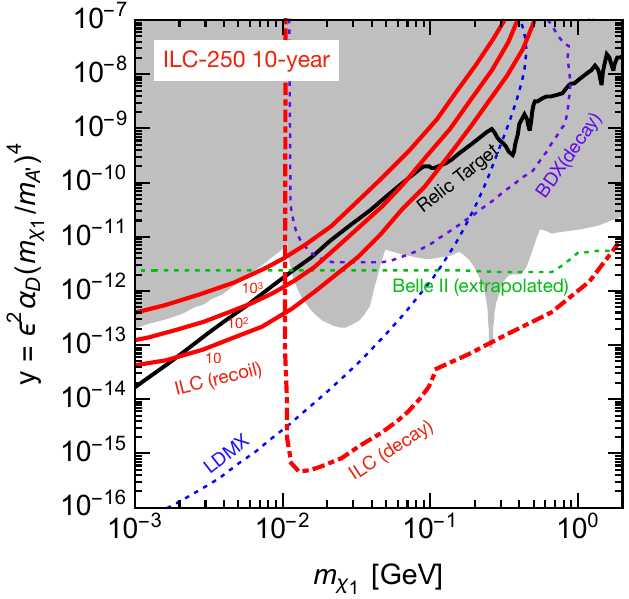}%
  \subcaption{electron beam dump}
  \end{subfigure}%
  \begin{subfigure}{0.49\textwidth}%
  \captionsetup{margin={25pt,0pt}}%
  \includegraphics[width=\textwidth]{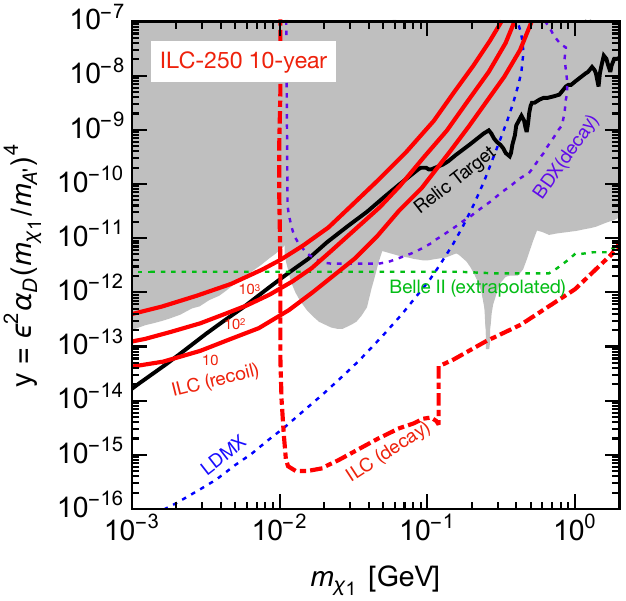}%
  \subcaption{positron beam dump}
  \end{subfigure}%
\caption{
    Projected sensitivity reach of a 10-year ILC-250 run in the pseudo-Dirac DM model with a large mass-splitting, $\Delta = 0.1m_{\chi_1}$.
  It is assumed that $\alpha_D=0.1$ and $m_{A'}=3 m_{\chi_1}$.
  The notation is similar to Fig.~\ref{fig:Dirac}; in addition, dot-dashed lines show the sensitivity of the ILC-BDX decay-signal search (95\% C.L. exclusion). Also shown are expected sensitivities of the BDX visible-decay search (purple-dashed lines)~\cite{BDX:2016akw,Izaguirre:2017bqb}, the LDMX (blue)~\cite{Berlin:2018bsc}, and  Belle~II (green)~\cite{Izaguirre:2015zva}. 
}
\label{fig:inelferm}
\end{figure}

Figure~\ref{fig:Dirac} summarizes our results for the small mass-splitting case, where we take the limit $\Delta\simeq0$ and fix $\alpha_D^{}=0.5$ and $m_{A'}^{}=3 m_{\chi}$.
The three red solid lines show the sensitivity of the ILC-BDX recoil-electron search with $\sqrt{s}=250\GeV$. The lines correspond to $10, 10^2$, and $10^3$ signal events with 10-year statistics.\footnote{%
  We also checked that the visible decay signals coming from $Z$ decays at the Giga-$Z$ program~\cite{Erler:2000jg} of the ILC are not significant for $m_{\chi}\lesssim 1$ GeV.
} The three lines illustrate the range of plausible scenarios for the level of reducible beam-induced backgrounds affecting the search, see Section~\ref{sec:BG}. 
The shaded regions are excluded by past experiments listed in Section~\ref{sec:cons}, and the solid black lines are the relic target for pseudo-Dirac DM.

Figure~\ref{fig:inelferm} shows the result for larger mass-splitting, where we fix $\Delta=0.1m_{\chi_1}$ as well as $\alpha_D^{}=0.1$ and $m_{A'}^{}=3m_{\chi_1}$.
Consequently, the visible-decay signal is expected for $m_{\chi_1}\gtrsim1\MeV$. It is noteworthy that the search for visible decays of $\chi_2$ provides a unique sensitivity for this model, well beyond that achievable at proposed dedicated future facilities such as LDMX. This strongly motivates the inclusion of a multi-layer tracker in the decay volume (see Fig.~\ref{fig:exp}) as a key component of the ILC-BDX design.

Below, we discuss the physics that determines the structure of the reach curves in Figs.~\ref{fig:Dirac} and~\ref{fig:inelferm}.

\subsubsection[Small mass-splitting]{Small mass-splitting: $\Delta < 2 m_{e}$}

Both pair-annihilation and bremsstrahlung are included as the source of signal events, where on-shell dark photons are produced and decay into a $\chi_2$-$\chi_1$ pair, and either $\chi_2$ or $\chi_1$ is detected as electron recoil.
\begin{itemize}
\item Pair-annihilation (recoil-electron) ---
The DM with mass less than $\simeq 10^{-2}$ GeV cannot be detected because of the threshold $E_{\rm min}=1~{\rm GeV}$.
This is because decays of the dark photon with mass less than $\sqrt{2 m_e E_{{\rm min}}}$ cannot contribute to signal events since the energy of the produced dark photon is $E^{\rm lab}_{A'}\simeq E_{e^+}\simeq m^2_{A'}/2m_e$.
Also, the DM with mass larger than $\simeq 10^{-1}$ GeV cannot be produced from decays of the dark photon because the dark photon mass cannot exceed $\sqrt{2 m_e E_{\rm beam}}$. 
In the positron beam dump for the DM mass of $\sqrt{2 m_e E_{\rm beam}}/3$, primary positron beam production dominates, and a peak structure arises.

\item Bremsstrahlung (recoil-electron) ---
For the DM mass smaller than $\sim 10^{-2}$ GeV, the number of the recoil-electron events is suppressed by the threshold $E_{\rm min}=1$ GeV.
Since the expected decay angle of the dark photon is $(\pi/2)\cdot(m_{A'}^{} /E_{A'})$, the dark photon energy has to satisfy $m_{A'}\cdot (\pi/2)\cdot (l_{\rm dump}+l_{\rm sh}+l_{\rm dec})/r_{\rm det}\lesssim E_{A'}$ to obtain sufficient angular acceptance.
However, the minimum energy of the dark photon is $E_{A'}\sim E_{\rm min}= 1$ GeV because of the threshold, and the signal events from the DM with mass smaller than $\sim 10^{-2}$ GeV are suppressed even if the angular acceptance holds. 
For the DM with mass larger than $0.1$ GeV, the angular acceptance becomes worse, and the sensitivity rapidly decreases.

\end{itemize}

\subsubsection[Large mass-splitting]{Large mass-splitting: $\Delta > 2m_e$}
The new feature in this case is the availability of the visible-decay signal. Here, we discuss the sensitivity reach for this channel, highlighting each of the DM decay processes.
The results of the electron recoil searches are similar to the case of the small mass-splitting.

\begin{itemize}
\item Pair-annihilation (visible-decay) ---
For $\Delta = 0.1 m_{\chi_1}$, the heavier DM with the mass smaller than $\simeq 10^{-2}$ GeV cannot decay, and the visible-decay signals do not arise.
Similar to the recoil-electron search, the DM with mass larger than $10^{-1}$ GeV cannot be produced because the dark photon mass cannot exceed $\sqrt{2 m_e E_{\rm beam}}$.
In the positron beam dump experiment, for $m_{\chi_1}=\sqrt{2m_e E_{\rm beam}}/3$, the peak structure arises due to the primary positron beam.

\item Bremsstrahlung (visible-decay) ---
Similar to the pair-annihilation process, for $\Delta = 0.1 m_{\chi_1}$, the heavier DM with the mass smaller than $\simeq 10^{-2}$ GeV cannot decay and produce the visible-decay signals. 
For $10^{-1}~{\rm GeV}\lesssim m_{\chi}$, the number of signal events rapidly decreases because the angular acceptance becomes worse.
\end{itemize}

\subsection{Scalar elastic DM}

\begin{figure}[t]
\centering
  \begin{subfigure}{0.49\textwidth}%
  \captionsetup{margin={25pt,0pt}}%
  \includegraphics[width=\textwidth]{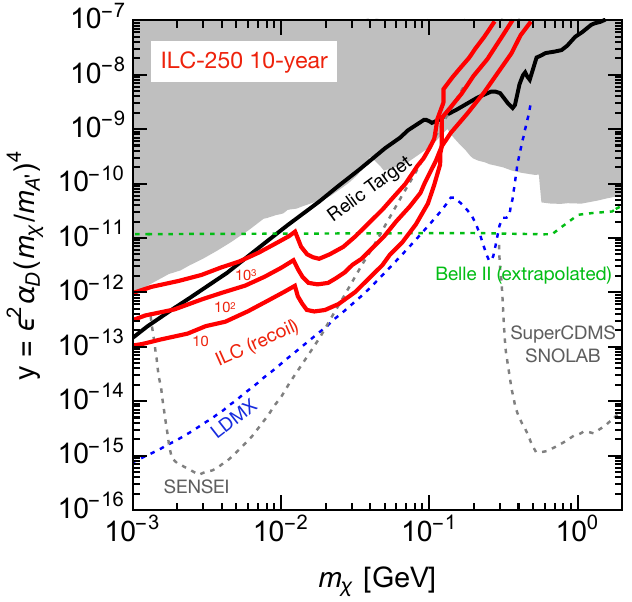}%
  \subcaption{electron beam dump}
  \end{subfigure}%
  \begin{subfigure}{0.49\textwidth}%
  \captionsetup{margin={25pt,0pt}}%
  \includegraphics[width=\textwidth]{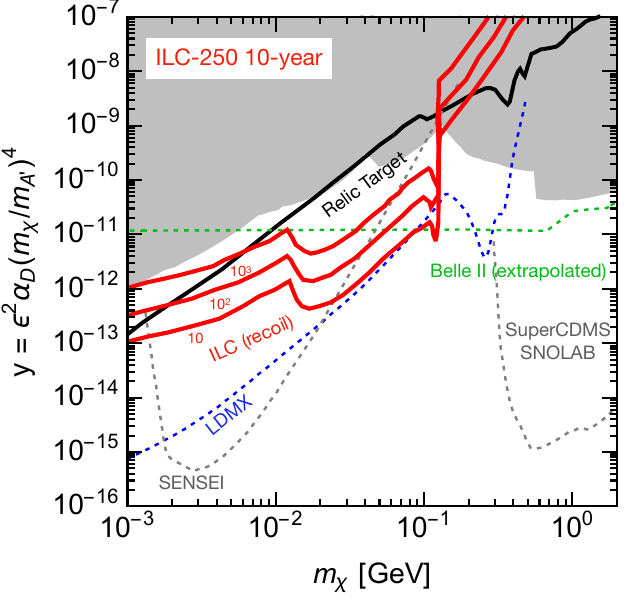}%
  \subcaption{positron beam dump}
  \end{subfigure}%
\caption{
The results for the scalar elastic DM.
It is assumed that $\alpha_D=0.5$ and $m_{A'}=3m_{\chi}$.
The notation is the same as in Fig.~\ref{fig:Dirac}; in addition, the gray lines show the expected sensitivities of SENSEI~\cite{Berlin:2018bsc} and SuperCDMS~\cite{Battaglieri:2017aum,Berlin:2018bsc}.
}
\label{fig:Scl}
\end{figure}

The model with a scalar elastic DM is described by a complex scalar field $\chi$ that is SM-singlet with unit U(1)$_D$-charge.
It is safe from CMB bounds since the DM annihilation is in $p$-wave and thus with a suppressed rate.
The DM current, originating in the kinetic term $|D_{\mu}\chi|^2$, is given by
\begin{equation}
    J^{\mu}_{\chi}=i (\chi^{\ast}\partial^{\mu}\chi -\chi\partial^{\mu}\chi^{\ast})~.
\end{equation}
The analysis results are shown in Fig.~\ref{fig:Scl} with the same notation as in Fig.~\ref{fig:Dirac}.
For the non-relativistic DM-SM fermion elastic scattering cross section, there is no velocity suppression, and the constraints from direct detection experiments are significant.

\subsection{Scalar inelastic DM}

\begin{figure}[t]
\centering
  \begin{subfigure}{0.49\textwidth}%
  \captionsetup{margin={25pt,0pt}}%
  \includegraphics[width=\textwidth]{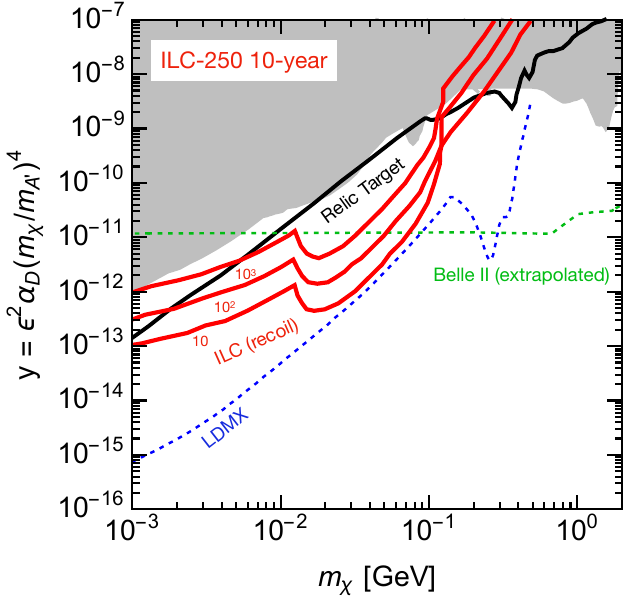}%
  \subcaption{electron beam dump}
  \end{subfigure}%
  \begin{subfigure}{0.49\textwidth}%
  \captionsetup{margin={25pt,0pt}}%
  \includegraphics[width=\textwidth]{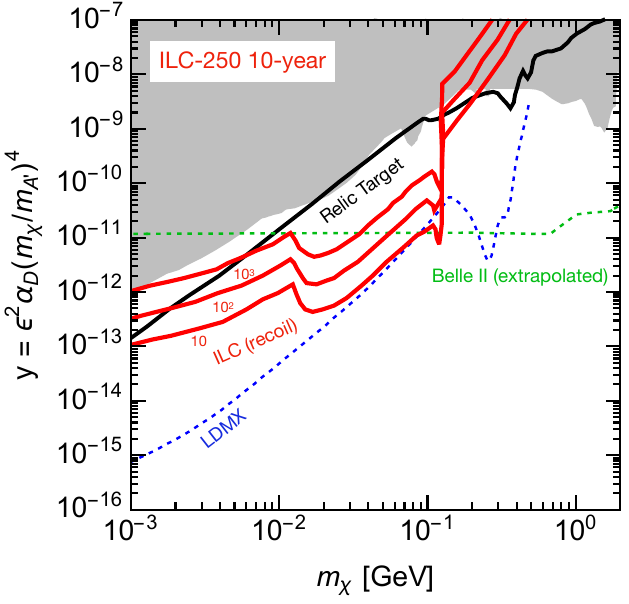}%
  \subcaption{positron beam dump}
  \end{subfigure}%
\caption{
The results for the scalar inelastic DM.
The prescription $\alpha_D^{}=0.5$, $m_{A'}^{}=3m_{\chi}$ is adopted.
Notation is the same as in Fig.~\ref{fig:Dirac}.
}
\label{fig:Sclinel}
\end{figure}

The scalar inelastic DM is characterized by the dark photon current
\begin{align}
    J^{\mu}_{\chi}=\chi_1 \partial^{\mu}\chi_2 -\chi_2 \partial^{\mu}\chi_1
\end{align}
involving two real scalar fields $\chi_{1,2}$ that are quasi-degenerate but have non-zero mass difference $\Delta=m_{\chi_2}-m_{\chi_1}>0$ after the spontaneous symmetry breaking of U(1)$_D$.\footnote{%
  The mass difference is realized by terms such as $\chi^2H^2$ after the electroweak (and U(1)$_D$) symmetry breaking, where $\chi$ is a complex scalar field yielding $\chi_{1,2}$, and $H$ is the Higgs doublet~\cite{Tucker-Smith:2001myb}.
See, e.g., Refs.~\cite{Cui:2009xq,Cheung:2009qd,Arina:2011cu,Okada:2019sbb} for models realizing scalar inelastic DM.}
We focus only on the small mass-splitting case since the large mass-splitting case of the scalar inelastic DM is similar to that of the pseudo-Dirac inelastic DM.
The results of our analysis are shown in Fig.~\ref{fig:Sclinel}, with the same notation as in Fig.~\ref{fig:Dirac}.

\subsection{Majorana DM}
\begin{figure}[t]
\centering
  \begin{subfigure}{0.49\textwidth}%
  \captionsetup{margin={25pt,0pt}}%
  \includegraphics[width=\textwidth]{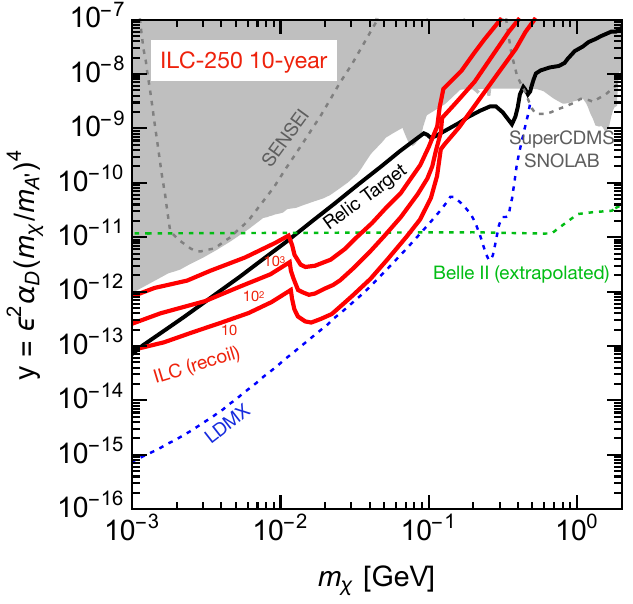}%
  \subcaption{electron beam dump}
  \end{subfigure}%
  \begin{subfigure}{0.49\textwidth}%
  \captionsetup{margin={25pt,0pt}}%
  \includegraphics[width=\textwidth]{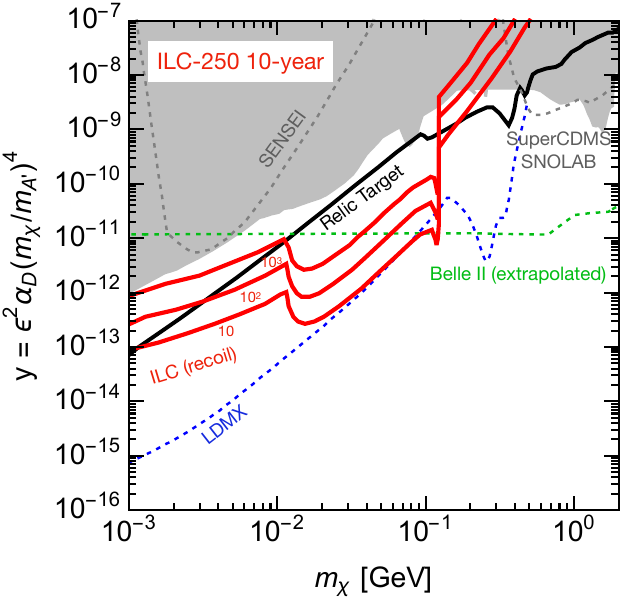}%
  \subcaption{positron beam dump}
  \end{subfigure}%
\caption{
  The results for the Majorana DM.
The prescription $\alpha_D^{}=0.5$, $m_{A'}^{}=3m_{\chi}$ is adopted.
Notation is the same as in Fig.~\ref{fig:Scl}.
}
\label{fig:Maj}
\end{figure}

For a Majorana fermion $\chi$, the DM current can be an axial-vector as follows:
\begin{align}
    J^{\mu}_{\chi}=\frac{1}{2}\bar{\chi}\gamma^{\mu}\gamma_5\chi.
\end{align}
Similar to the scalar DM, the pair-annihilation rate of the DM into SM particles is suppressed by the DM velocity, and a sub-GeV Majorana fermion DM is safe from the CMB constraints. The results of our analysis for the Majorana DM are shown in Fig.~\ref{fig:Maj}, with the same notation as in the scalar elastic DM case (Fig.~\ref{fig:Scl}).

\subsection{Existing constraints and projected sensitivities at other experiments}
\label{sec:cons}

Searches for sub-GeV DM are pursued by accelerator-based experiments and direct detection experiments.
In the accelerator-based experiments, the DM particles may be produced by beam-beam or beam-target collisions, and the searches are performed either by detecting their scattering off of the SM particles in the detectors, the visible decays of the excited DM states, or the missing energy/momentum events.  
In the direct detection experiments, the DM particles from the halo are searched for by detecting their scattering off of the SM particles in the detectors.
We list the existing constraints included in Figs.~\ref{fig:Dirac}--\ref{fig:Maj} as follows:

\begin{itemize}

\item E137 electron beam dump experiment --- 
The light DM may be produced at the SLAC electron beam dump by injection of 20 GeV electron beams~\cite{Batell:2014mga}.
The produced DM particles may be detected by their scattering on electrons in the aluminum and plastic scintillator 400 m downstream from the beam dump.
The visible decays of the heavier dark sector particles are also detectable at the scintillator~\cite{Izaguirre:2017bqb}. 

\item NA64 --- The NA64 experiment~\cite{Andreev:2021fzd} at CERN constrains the light DM by analyzing data from interaction of $2.84\times 10^{11}$ electrons at 100 GeV with an active thick target, i.e., electromagnetic calorimeter, and searching for missing-energy events.
The DM may be produced in the active target by decays of dark photons generated via the bremsstrahlung process and secondary positron annihilation with atomic electrons.

\item LSND proton beam dump experiment --- 
The light DM may also be produced in the Los Alamos LSND proton beam dump experiment by injection of proton beams with kinetic energy of 800 MeV~\cite{deNiverville:2011it,Izaguirre:2017bqb,Berlin:2018bsc,Berlin:2018pwiSea}.
The DM particles are produced dominantly by $\pi^0$ decays at the fixed target, and DM-electron scattering or visible decay events are detected at the downstream detector.

\item MiniBooNE --- A search for the light DM produced from the Fermilab 8 GeV Booster protons with a steel beam dump~\cite{MiniBooNEDM:2018cxm} was conducted by the MiniBooNE-DM collaborations by using data from $1.86\times 10^{20}$ protons on target (POT).
To reduce neutrino backgrounds from charged mesons, off-target running was conducted, and the primary proton beam was steered above the thin beryllium target and directed into the thick steel absorber.
Charged mesons produced in the thick target are absorbed or decay at rest, and the neutrino flux is reduced.
The DM may be produced by decays of $\pi^0$ and $\eta$ mesons and proton bremsstrahlung plus vector-messon mixing.
The produced DM may be detected by elastic scattering off nucleons, inelastic neutral pion production, and elastic scattering-off electrons in the MiniBooNE detector, consisting of 818 tons of mineral oil and placed 490 m downstream of the beam dump.

\item COHERENT CsI --- The COHERENT experiment~\cite{COHERENT:2021pvd,Akimov:2022oyb} in Neutrino Alley at the Spallation Neutron Source (SNS) Oak Ridge National Laboratory has made the first measurement of coherent elastic neutrino-nucleus scattering (CE$\nu$NS) predicted by the SM.
The 1.4 MW beam of 1 GeV protons is directed onto a thick and dense liquid mercury target at the rate of $10^{16}$ POT per second. 
The light DM may be produced via decays of neutral mesons such as $\pi^0$ and $\eta$, $\pi^-$ absorption process by the so-called Panofsky process $\pi^-+p\to n+A'$, and the dark photon bremsstrahlung process.
The coherent condition, where the neutrino and DM scatters off all nucleons of the nucleus, is satisfied because of low beam energy and enhances the cross-section of the detection processes.
The neutrino flux is isotropic, and detectors are placed at off-axis angles from the beam.
Although about 99\% of $\pi^-$ produced are captured in the thick target, the majority of $\pi^+$ stops and yields prompt $\nu_{\mu}$ flux, and the majority of $\mu^+$ yields delayed $\nu_e$ and $\bar{\nu}_{\mu}$ fluxes.
Adopting a timing cut, the neutrino background events lying in delayed timing bins are rejected.
The initial COHERENT data using a 14.6kg CsI[Na] scintillator detector placed at a distance of 19.3 m from the target set limits on the DM.

\item CCM120 --- One of the main missions of the Coherent CAPTAIN-Mills (CCM) detector~\cite{CCM:2021leg} located at the Los Alamos Neutron Science Center at Los Alamos National Laboratory is the light DM search.
800 MeV bunched proton beam is directed onto a thick tungsten target, and then the DM may be produced from decays of $\pi^0$.
Each bunch consists of $\sim 3.1\times 10^{13}$ protons with a repetition rate of $20$ Hz.
The CCM120 detector containing 120 PMTs is a 10-ton liquid argon detector placed at a distance of 20 m from the target.
The DM may be detected through coherent elastic DM-neucleus scattering, quasi-elastic nucleon scattering, and elastic electron scattering.
The neutrino backgrounds are eliminated by energy and timing selection.
An engineering run with 1.5 months of data set limits on the DM.

\item { CHARM proton beam dump experiment --- The DM particles may be produced at a proton beam dump experiment operating at the 400 GeV CERN SPS. Decays of mesons such as $\pi^0$ and $\eta$ and proton bremsstrahlung processes yield the heavier dark sector particles, which can be detected via the visible decay signature~\cite{Tsai:2019buq}.}

\item {$\nu$-calorimeter I neutrino experiment --- The visible-decay search for the heavier dark sector particles was conducted in the neutrino experiment. Similar to the CHARM experiment, the visible-decay search is sensitive to the meson decays and the bremsstrah-lung processes~\cite{Tsai:2019buq}.}

\item $\gamma$+missing energy search --- 
The DM may be produced at B-factories by process $e^+ e^-\to \gamma +{A'}^{(\ast)}\to \gamma\chi\bar{\chi}$.
The mono-photon and missing-energy events at the BaBar set bounds on the DM models~\cite{Izaguirre:2013uxa,Essig:2013vha,Berlin:2018bsc}. 

\item  Direct detection experiments --- 
The DM-SM non-relativistic scattering may be detected in direct detection experiments, such as XENON-1T, SENSEI, SuperCDMS and EDELWEISS~\cite{XENON:2019zpr,XENON:2021qze,SENSEI:2020dpa,SuperCDMS:2020aus,SuperCDMS:2023sql,EDELWEISS:2019vjv,EDELWEISS:2022ktt}.
The bounds from these direct detection experiments are included in the shaded gray region of Fig.~\ref{fig:Scl}.
For the DM-nuclear scattering process, the SuperCDMS experiment~\cite{SuperCDMS:2020aus} is sensitive to $m_{\chi}\sim 1$ GeV because of the low nuclear recoil energy threshold. 
Also, the DM-nuclear scattering searches, e.g., XENON-1T~\cite{XENON:2019zpr}, EDELWEISS~\cite{EDELWEISS:2019vjv,EDELWEISS:2022ktt} and SuperCDMS~\cite{SuperCDMS:2023sql}, are sensitive to sub-GeV DM through inelastic scattering channels: bremsstrahlung radiation and the Migdal effect.
The sub-GeV DMs with MeV to GeV mass may also be explored using the DM-electron scattering searches, e.g., XENON-1T~\cite{XENON:2021qze} and SENSEI~\cite{SENSEI:2020dpa}.

\end{itemize}

\begin{figure}[p]
\centering
  \captionsetup{margin={25pt,0pt}}%
  \includegraphics[width=0.99\textwidth]{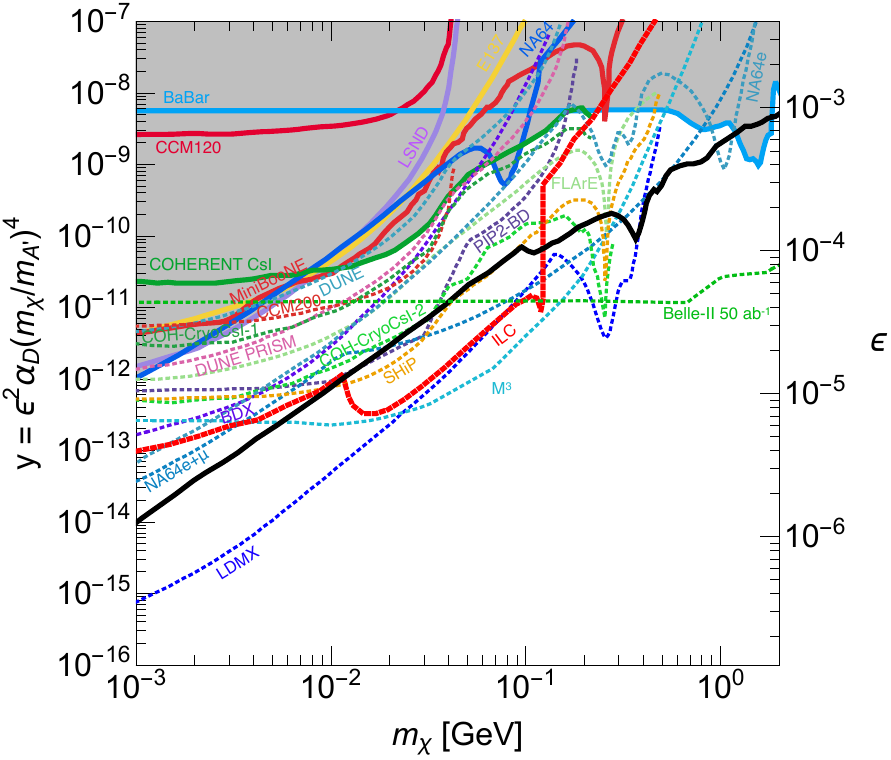}%
\caption{
   Constraints on the pseudo-Dirac DM with small mass-splitting (shaded gray), projected sensitivities of future projects (dashed lines), and the results of a 10-year run of ILC-250 positron BDX, assuming $\alpha_D=0.5$ and $m_{A'}=3 m_{\chi}$ (red solid line).
   %
}
\label{fig:pDirac}
\end{figure}
\begin{figure}[p]
\centering
  \captionsetup{margin={25pt,0pt}}%
  \includegraphics[width=0.99\textwidth]{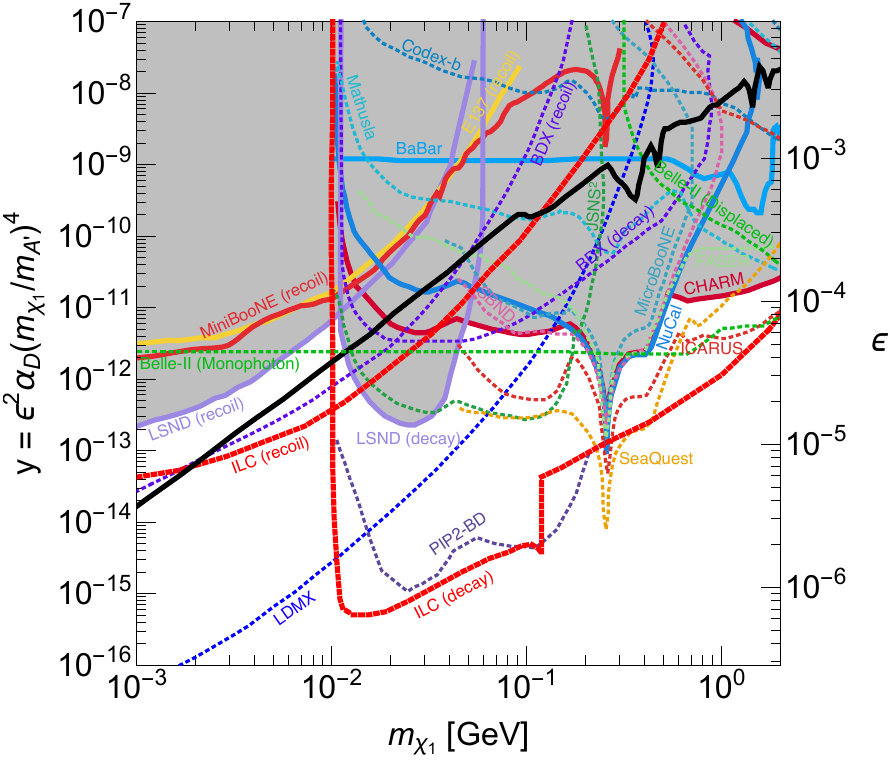}%
\caption{
      Constraints on the pseudo-Dirac DM with large mass-splitting (shaded gray), projected sensitivities of future projects (dashed lines), and the results of a 10-year run of ILC-250 positron BDX, assuming $\Delta=0.1 m_{\chi_1}$, $\alpha_D=0.1$ and $m_{A'}=3 m_{\chi_1}$ (red solid line).
   See Section~\ref{sec:cons} for the details of the constraints and future projects.
}
\label{fig:inDirac}
\end{figure}

As representative future accelerator-based searches for light DM, we included projected sensitivities at Belle II~\cite{Belle-II:2018jsg,Duerr:2019dmv}, BDX~\cite{BDX:2016akw}, and LDMX~\cite{Berlin:2018bsc,Akesson:2022vza} in Figs.~\ref{fig:Dirac}-\ref{fig:Maj} to compare with the sensitivity reaches of the ILC-BDX.
Many other future experiments have been discussed in the literature. 
For completeness, Figs.~\ref{fig:pDirac} and \ref{fig:inDirac} compare the ILC-BDX sensitivity reaches to a broader set of future experiments. The following experiments are included:

\begin{itemize}
    \item LDMX --- The Light Dark Matter eXperiment (LDMX)~\cite{Berlin:2018bsc,Akesson:2022vza} at the SLAC is an electron-beam fixed-target missing-momentum experiment that explores the light DM in the sub-GeV range.
    Once a single electron collides with a thin tungsten target, the final state is reconstructed by analyzing the downstream apparatus, which includes a charged particle tracker and calorimeters.
    DM signals may be produced via the dark photon bremsstrahlung process and invisible decays of vector mesons produced by a hard photon.
    The LDMX is designed for $4\times 10^{14}$ electrons on target (EOT) at a 4 GeV beam energy (Phase 1) and is upgraded to $10^{16}$ EOT at an 8 GeV beam energy (Phase 2). 
    The sensitivity of Phase 2 is shown in the figures.
    \item M$^3$ --- The muon missing momentum (M$^3$)~\cite{Kahn:2018cqs} at the Fermilab is a muon-beam reconfiguration of the proposed LDMX experiment, which would be conducted at a 15 GeV muon beam with $10^{10}$ muons on target (Phase 1) and $10^{13}$ muons on target (Phase 2).
    The sensitivity of Phase 2 is shown in Fig.~\ref{fig:pDirac}.
    \item BDX --- The Beam-Dump eXperiment (BDX)~\cite{BDX:2016akw} at the Jefferson Lab is conducted with $10^{22}$ EOT at 11 GeV.
    DM particles are detected through their scattering on electrons within a calorimeter approximately one cubic meter in size, situated 20 m away from the beam dump.
    \item NA64e --- The future run of the NA64 experiment will be conducted with $5\times 10^{12}$ electrons at 100 GeV.
    In addition to the bremsstrahlung process~\cite{Gninenko:2020hbd} and secondary positron annihilation with atomic electrons, the DM may be produced by invisible decays of vector mesons produced by a hard photon~\cite{Schuster:2021mlr}.
    \item NA64$\mu$ --- The NA64$\mu$ experiment~\cite{Gninenko:2020hbd} at CERN may constrain the light DM by searching missing-energy events through the interaction of muon beams with an active thick target.
    The DM may be produced in the active target by decays of dark photons via the bremsstrahlung process.
    The sensitivity of the NA64$\mu$ with $2\times 10^{13}$ muons at 100 GeV on an active thick target is shown in Fig.~\ref{fig:pDirac}.
    \item Belle II --- The $\gamma$+missing energy search can be conducted at the Belle-II~\cite{Belle-II:2018jsg}.
    The sensitivity of the Belle-II with an integrated luminosity of 50 ab$^{-1}$ is shown in the figures.
    In the inelastic DM model with large mass-splitting, if the $\chi_2$ decays outside the detector, the final state appears identical to the process of $e^+ + e^- \to \gamma + A', A'\to$ invisible.
    The same signature arises if the decay vertex of $\chi_2$ is located within the detector, but the energy of the decay products is insufficient to be detected.
    Alternatively, it is possible to detect the decay products of $\chi_2$ and reconstruct a pair of displaced leptons or hadrons (such as pions and kaons) in conjunction with a single photon.
    The sensitivity of the displaced signatures at the Belle-II~\cite{Duerr:2019dmv} with an integrated luminosity of 50 ab$^{-1}$ is shown in Fig.~\ref{fig:inDirac}.
    \item COH-CryoCsI-1 and COH-CryoCsI-2 --- The COHERENT experiment~\cite{Akimov:2022oyb} will conduct cryogenic operations to eliminate backgrounds emitted from PMTs.
    A 10 kg CsI (COH-CryoCsI-1) and a 700 kg CsI (COH-CryoCsI-2) will be operated at 40 K and placed at a distance of 20 m from the target.
    \item PIP2-BD --- The Proton Improvement Project II (PIP-II) is the first phase of a major transformation of the accelerator facilities at Fermilab to prepare the DUNE.
    The PIP2-BD~\cite{Toups:2022yxs} will be conducted with a graphite target and a 100-tonne scintillation-only liquid argon detector placed 18 m from the target.
    The sensitivity of PIP2-BD with a 1.2 GeV proton bunch consisting of $1.2\times 10^{12}$ protons and a repetition rate of 100 Hz for five years is shown in Fig.~\ref{fig:pDirac}.
    \item DUNE --- A near detector of the Deep Underground Neutrino Experiment (DUNE)~\cite{DUNE:2020lwj} will be located on-site at Fermilab and placed on the beam axis 574 m from the graphite target.
    120 GeV proton beam is directed onto the target with $1.1 \times 10^{21}$ protons on target collected per year.
    \item DUNE-PRISM --- The DUNE-PRISM~\cite{DeRomeri:2019kic} is the concept to move the DUNE near detector up to $\sim 36$\,m transverse to the beam axis to reduce background induced by neutrinos.
    \item CCM200 --- The CCM experiment will be upgraded with the new CCM200 detector consisting of 200 PMTs and filtered liquid argon~\cite{CCM:2021leg}.
    The sensitivity of the CCM200 for a three-year run collecting $2.25\times 10^{22}$ POT is shown in Fig.~\ref{fig:pDirac}.
    \item SHiP --- The SHiP experiment at CERN may constrain the light DM by the interaction of DMs with neutrino detectors.
    After 400 GeV proton beam collisions with a target, the DM may be produced by decays of $\pi^0$ and $\eta$, the bremsstrahlung process, the Drell-Yan-like production, and the associated production with QCD radiation.
    The DMs may be detected by elastic scattering off electrons ($\chi e^-\to \chi e^-$) and off protons ($\chi p\to \chi p$).
    In estimating the projected sensitivity of SHiP~\cite{SHiP:2020noy}, 5-years of running corresponding to $2\times 10^{20}$ POT is assumed.
    \item FASER --- The ForwArd Search ExpeRiment (FASER) at LHC is an experiment designed to search for light long-lived particles in the very forward region.
    The FASER is placed along the beam collision axis, 480 m downstream from the ATLAS IP.
    In estimating the projected sensitivity of FASER~\cite{Berlin:2018jbm}, an integrated luminosity of 3 ab$^{-1}$ is assumed.
    \item FLArE --- The Forward Liquid Argon Experiment (FLArE) detector~\cite{Kling:2022ykt} is planed to be placed 620\,m downstream from the ATLAS interaction point.
    DM particles are produced through the bremsstrahlung proces and meson rare decays at the ATLAS collision point.
    The sensitivity in the High Luminosity LHC run with an integrated luminosity of $3 {\rm ab}^{-1}$ and 14 TeV proton-proton collision energy is shown in Fig.~\ref{fig:pDirac}.
    \item Mathusla --- The MAssive Timing Hodoscope for Ultra-Stable neutraL pArticles (MATHUSLA) at LHC is planned to be located on the surface near ATLAS or CMS.
    The detector consists of a 200 m $\times$ 200 m $\times$ 20 m decay volume, positioned $\sim$ 100 m downstream from the IP and $\sim$ 100 m above the LHC beam.
    %
    %
    The decay volume is covered with a scintillating layer to veto incoming charged particles like high-energy muons from the IP.
    In estimating the projected sensitivity of Mathusla~\cite{Berlin:2018jbm}, $\chi_2$ decay in the decay volume is the signal event, and an integrated luminosity of 3 ab$^{-1}$ is assumed. 
    \item Codex-b --- The COmpact Detector for EXotics at LHCb (CODEX-b) has the decay volume with dimensions of 10 m $\times$ 10 m $\times$ 10 m and is to be positioned $\sim$ 5 m downstream from the IP at a transverse distance of $\sim$ 26 m.
    The Codex-b detector is proposed to be constructed in the LHCb cavern.
    In estimating the projected sensitivity of Codex-b~\cite{Berlin:2018jbm}, $\chi_2$ decay in the decay volume is the signal event, and LHCb's ultimate luminosity of 300 fb$^{-1}$ is assumed. 
    \item SeaQuest --- The SeaQuest experiment~\cite{Berlin:2018pwi} at the Fermilab is a beam dump experiment operated with a 120 GeV proton beam of $\sim 1.44\times 10^{18}$ POT (Phase 1) and $10^{20}$ POT (Phase 2).
    %
    %
    The proton beam is dumped onto a thin nuclear target and thick iron magnet. 
    The DM may be produced through the dark photons, which are generated by decays of mesons such as $\pi^0, \eta,\omega$, and $\eta'$, proton bremsstrahlung, and the Drell-Yan production.
    \item JSNS$^2$ --- The J-PARC Sterile Neutrino Search at the J-PARC Spallation Neutron Source (JSNS$^2$) experiment with 3 GeV kinetic energy protons may be sensitive to the visible decay of $\chi_2$ and DM-electron scattering events.
    Neutral mesons such as $\pi^0$ and $\eta$ are produced by proton beam collisions with a mercury target, and the DM may be produced via decays of such mesons.
    The JSNS$^2$ liquid scintillator-based detector is located 24 m downstream from the target.
    The sensitivity of JSNS$^2$~\cite{Jordan:2018gcd} for a one-year run collecting 3.8$\times 10^{22}$ POT/year is shown in Fig.~\ref{fig:inDirac}. 
    \item MicroBooNE --- The Fermilab Short-Baseline Neutrino program with the liquid argon time projection chamber (LArTPC) detector MicroBooNE installed along the Booster Neutrino Beam (BNB) may be sensitive to the visible decay of $\chi_2$ and DM-nucleon scattering events~\cite{Batell:2021ooj}.
    The BNB is produced by removing protons from the Booster accelerator at 8 GeV kinetic energy and colliding them with a beryllium target using $6.6\times 10^{20}$ POT.
    The detector is located on the beam axis 470 m from the BNB target.
    The dark photons may be produced through three mechanisms: decays of pseudoscalar meson $\pi^0,\eta,\eta' \to \gamma A'$, proton bremsstrahlung, and the Drell-Yan process. 
    \item SBND --- The SBND at the Fermilab is the LArTPC detector installed along the BNB using 8 GeV kinetic energy protons with $6.6\times 10^{20}$ POT~\cite{Batell:2021ooj}. 
    The detector is located on the beam axis 110 m away from the BNB target.
    \item ICARUS --- The ICARUS at the Fermilab is the LArTPC detector installed along the BNB~\cite{Batell:2021ooj}.
    The detector is located on the beam axis 600 m from the BNB target.
    Also, the ICARUS detector is positioned within a 6-degree off-axis angle from the Neutrinos at the Main Injector (NuMI) beam, which is created by 120 GeV protons colliding with a graphite target at $7.7\times 10^{21}$ POT.
\end{itemize}
As the future direct detection experiments, in Fig.~\ref{fig:Scl}, we included projected sensitivities of SuperCDMS~\cite{SuperCDMS:2022kse} and SENSEI~\cite{SENSEI:2020dpa,Battaglieri:2017aum} at SNOLAB.
The sensitivity of the SENSEI SNOLAB is obtained in Ref.~\cite{Battaglieri:2017aum} for the planned construction of the 100-gram SENSEI experiment~\cite{SENSEI:2020dpa} of exposure over 1-year.

\section{Summary}
\label{sec:Summary}

In this paper, we explored the potential of a search for sub-GeV dark matter at the ILC beam dumps using the ILC-BDX setup. Previous studies of beam dump experiments at the ILC 
focused on long-lived BSM particles decaying into visible SM final states. In this study, on the contrary, we consider DM models where a dark photon connects the stable DM to the SM sector, and focus on DM detection.
Five DM models, i.e., pseudo-Dirac DM with small and large mass-splitting, scalar elastic and inelastic DM, and Majorana DM are considered as reference scenarios.
At the ILC, electrons and positrons in electromagnetic shower are produced by the beam injection into the beam dumps. These particles produce dark photons through the bremsstrahlung and pair-annihilation processes, which in turn decay to pairs of DM particles. (Pair-production of DM via virtual dark photon exchange is also possible,) 
Produced DM particles propagate over a long distance (of order 100 m) to the detector and elastically scatter with electrons inside the detector, and then the recoil electrons are detected as a signature of the DM production.
For the inelastic DM models, there is another signal produced by a visible decay of the heavier dark state into the lighter one and SM particles. The ILC-BDX setup includes a multi-layer tracker designed to search for this visible-decay signal. 

In the DM search at the ILC-BDX, there are two kinds of backgrounds, i.e., the beam-induced and beam-unrelated BGs. The beam-induced BG is due to neutrinos mainly produced by meson and lepton decays in the beam dumps. Elastic neutrino-electron scattering in the detector is nearly indistinguishable from DM-induced electron recoil events, providing irreducible BGs to this search. Inelastic neutrino-nucleus interactions can also mimic electron-recoil and visible-decay signals, though these backgrounds are potentially reducible. We evaluated the neutrino-induced irreducible and reducible BGs by using the neutrino fluxes at the detector calculated with Monte Carlo simulation. We also estimated the beam-unrelated BG, which comes predominantly from cosmic-ray muons. This BG component can be significantly suppressed by the deep underground location of the detector and beam-coincidence time window.

We evaluated the number of DM particles produced at the ILC electron and positron beam dumps and the rate of expected signal events, i.e., electron-DM recoils and visible-decays at the ILC-BDX, in our reference models. The predicted signal and background rates were then used to estimate the sensitivity reach of the ILC-BDX experiment with a data set corresponding to a 10-year ILC run at $\sqrt{s}=250$~GeV. The results for each DM model are shown in Figs.~\ref{fig:Dirac}-\ref{fig:Maj}.
We found that in all five reference models considered here, the ILC-BDX experiment has sensitivity to regions of model parameter space well beyond the existing constraints. In many cases, the ILC-BDX can conclusively probe the region of parameter space where the DM has the thermal abundance consistent with observations (the ``relic target"). We find that while electron and positron beam dumps have similar performances, the sensitivity of the latter is somewhat better because the primary positron beam contributes to the DM production through pair-annihilation on electrons in the dump. The sensitivity of the ILC-BDX is particularly impressive in models where the visible-decay signal is available, such as the pseudo-Dirac or scalar inelastic DM models with mass-splitting $\Delta > 2m_e$. In such models, the reach of the ILC-BDX significantly exceeds that of even the most ambitious proposed dedicated searches for sub-GeV DM, such as LDMX.   

In the discussion of future searches for sub-GeV dark matter, it is important to note the complementarity between the electron-recoil technique used by ILC-BDX and the missing energy/momentum technique used by experiments such as LDMX. The signal rate at missing energy experiments is simply proportional to the total cross section of DM production by the beam electron interactions with the target material. On the other hand, the signal rate at electron-recoil experiments is proportional to the {\em product} of this production cross section and that of the elastic scattering of DM on target electrons. The relationship between the production and elastic scattering cross sections is model-dependent. Thus, analyzing data from both types of experiments will provide an opportunity to identify the underlying DM model, if a signal is observed with one or both approaches. This is an exciting possibility, and it strongly motivates pursuing both approaches in parallel in the future.    

In our evaluation of the acceptance of the signal events, we adopted some simplifying assumptions, for instance, the approximated angular distribution of the heavy DM state in Eq.~\eqref{eq:ang-distribution} and radial deviations in Eqs.~\eqref{eq:radial-dev} and \eqref{eq:radial-dev-recoil}. Likewise, several simplifying assumptions were made in the estimates of the background rates; see Section~\ref{sec:BG}. As a next step, it would be important to design a realistic Monte Carlo model of the ILC-BDX detector, and use it to provide a more precise evaluation of both signal and background rates. Another important direction for future work is to design strategies for handling the reducible beam-induced BG, such as additional cuts or a veto system, and to evaluate such strategies quantitatively using Monte Carlo simulations.

\section*{Acknowledgements}
We are grateful to the members of the ILC Task Force on fixed-target experiments and dark sectors, especially Claude Vallee, for discussions that initiated this work. KA is supported by JSPS KAKENHI Grant Number JP18H01210, JP21K20365, and MEXT KAKENHI Grant Number JP18H05543. YS is supported by JSPS KAKENHI JP21H05466. MP is supported by the NSF grant PHY-2014071.

\appendix
\section{Dark photon production cross sections}
\label{app:prodcross}

In this appendix, we list the cross sections of dark photon production via pair-annihilation and bremsstrahlung.

\subsection{Pair-annihilation production}

The cross section of the resonant annihilation process is given by~\cite{Marsicano:2018krp}
\begin{align}
    \sigma(e^+ e^-\to A')=\frac{12\pi}{m^2_{A'}}\frac{\Gamma^2_{A'}/4}{(\sqrt{s}-m_{A'})^2+\Gamma^2_{A'}/4},
\end{align}
where $s$ is the center-of-mass energy squared and $\Gamma_{A'}$ is the total decay width of the dark photon.
In this work, $\Gamma_{A'}$ is assumed to be small enough that the narrow-width approximation can be used:
\begin{align}
    \sigma (e^+ e^- \to A')\simeq \frac{2\pi^2 \alpha \epsilon^2}{m_e}\mathop{\delta}\left(E_i -\frac{m^2_{A'}}{2 m_e}+m_e\right),
\end{align}
where $\delta (x)$ denotes the Dirac delta function.

\subsection{Bremsstrahlung production}
The differential cross section of the bremsstrahlung process under the Weizs\"acker-Williams approximation~\cite{vonWeizsacker:1934nji,Williams:1935dka,Kim:1973he,Tsai:1986tx} is given by~\cite{Liu:2017htz,Bjorken:2009mm}
\begin{align}
    \frac{\dd^2 \sigma (i {\rm N}\to i A' {\rm N})}{\dd E_{A'}\,\dd\theta_{A'}}=\epsilon^2 \alpha^3 \frac{1}{E_i}\sin\theta_{A'}\frac{\sqrt{E_{A'}^2-m_{A'}^2}}{\sqrt{E_i^2-m_e^2}}\frac{1}{1-x}\frac{\mathcal{A}^{2\to 2}_{t=t_{\rm min}^{\rm WW}}}{2 t_{\rm min}^{\rm WW}}\chi,
\end{align}
where $x=E_{A'}/E_i$, $t_{\rm min}^{\rm WW}=\tilde{s}^2/4 E_i^2$, $\tilde{s}=-\tilde{u}/(1-x)$, and
\begin{align}
\tilde{u}=-x E_i^2 \theta_{A'}^2 -m^2_{A'}\frac{1-x}{x}-m_e^2 x.
\end{align}
The effective flux of photons, $\chi$, is given by
\begin{align}
\chi&=\int_{t_{\rm min}}^{t_{\rm max}} \dd t\, \frac{t-t_{\rm min}}{t^2} G_2(t)
\notag\\
\begin{split}
    &=Z^2 \left(1-\frac{b}{a}\right)^{-2}\bigg[
    -\frac{(a+b+2ab t_{\rm max})(t_{\rm max}-t_{\rm min})}{(1+a t_{\rm max})(1+b t_{\rm max})}
    \\
    &\qquad\qquad\qquad
    +\frac{a+b+2ab t_{\rm min}}{a-b} \ln \left(\frac{1+a t_{\rm max}}{1+b t_{\rm min}}\,\frac{1+b t_{\rm min}}{1+a t_{\rm min}}\right)
    \bigg]
\end{split}
\end{align}
with $t_{\rm min}={m^4_{A'}}/{4 E_i^2}$, $t_{\rm max}=m^2_{A'}+m_e^2$, and
\begin{align}
&G_2(t)=\left[ 
\left(\frac{at}{1+at}\right)
\left(\frac{1}{1+bt} \right)Z\right]^2,\quad
a=\frac{112^2 Z^{-2/3}}{m^2_e},\quad
b=\frac{1}{0.164~{\rm GeV}^2 A^{-2/3}},
\end{align}
where $Z$ is the atomic number and $A$ is the mass number of the target atom.
The amplitude under the the Weizs\"acker-Williams approximation is given by
\begin{align}
    \mathcal{A}^{2\to 2}_{t=t_{\rm min}^{\rm WW}}=2 \frac{2-2x+x^2}{1-x}+4 (m^2_{A'}+2 m_e^2) \frac{\tilde{u} x + m^2_{A'} (1-x)+m^2_e x^2}{\tilde{u}^2}.
\end{align}

\section{DM-electron recoil cross sections}
\label{app:recoilcross}
For the inelastic-fermion DM model and the scalar DM model, the cross section of the DM-electron recoil process is given by~\cite{Kim:2016zjx,Giudice:2017zke},
\begin{align}
    \frac{\dd \sigma (\chi_1 e\to \chi_2 e)}{\dd E_e}=
    \frac{m_e}{8\pi \lambda(s,m_e^2,m_{\chi_1}^2)}|\overline{\mathcal{M}}|^2,
\end{align}
where $E_e$ is the energy of the recoil electron, $\lambda(x,y,z)=(x-y-z)^2-4 yz$, $E_e=E_{\chi_1}+m_e-E_{\chi_2}$, and $s=m_{\chi_1}^2 +m_e^2+2 m_e E_{\chi_1}$.
The squared matrix elements are given by
\begin{equation}
\begin{split}
  |\overline{\mathcal{M}}|^2
  &=
  \frac{8m_e (\epsilon e g_D^{})^2}{[2m_e (E_{\chi_2}-E_{\chi_1})-m_{A'}^2]^2}
  \biggl[
    m_e (E_{\chi_1}^2+E_{\chi_2}^2)
    \\&\qquad\qquad
    -\frac{\Delta^2}{2} (E_{\chi_2}-E_{\chi_1}+m_e)+m_e^2 (E_{\chi_2}-E_{\chi_1}) +m_{\chi_1}^2 E_{\chi_2} -m_{\chi_2}^2 E_{\chi_1}
  \biggr]
\end{split}
\end{equation}
for the pseudo-Dirac inelastic DM model and
\begin{equation}
  |\overline{\mathcal{M}}|^2=\frac{8m_e (\epsilon e g_D^{})^2}{[2m_e (E_{\chi_2}-E_{\chi_1})-m_{A'}^2]^2}
  \biggl[
    2m_e E_{\chi_1} E_{\chi_2} +m_{\chi_1}^2 E_{\chi_2} -m_{\chi_2}^2 E_{\chi_1}
  \biggr]
\end{equation}
for the scalar inelastic DM model,
where $E_{\chi_i}$ $(i=1,2)$ is the energy of $\chi_i$ measured in the laboratory frame.
The kinematically allowed maximum (minimum) recoil energy $E_e^+$ ($E_e^-$) is given by~\cite{Kim:2016zjx}
\begin{equation}
    E^{\pm}_e =\frac{s +m_e^2 -m_{\chi_2}^2}{2\sqrt{s}}\frac{E_{\chi_1} +m_e}{\sqrt{s}}\pm \frac{\sqrt{\lambda (s,m_e^2,m_{\chi_2}^2)}}{2\sqrt{s}}\frac{p_{\chi_1}}{\sqrt{s}},\label{eq:minrecoil}
\end{equation}
where $p_{\chi_1}=\sqrt{E_{\chi_1}^2-m_{\chi_1}^2}$.

The recoil cross section for the other models are given by~\cite{Kim:2016zjx,Giudice:2017zke,Batell:2014mga,Batell:2021blf}
\begin{equation}
    \frac{\dd \sigma (\chi e\to \chi e)}{\dd E_e}
    =
      4\pi \epsilon^2 \alpha \alpha_D
    \frac{
      2m_e E^2_{\chi}-(2m_e E_{\chi}-m_e E_e +m_{\chi}^2+2m_e^2)(E_e-m_e)
    }
    {(E_{\chi}^2-m_{\chi}^2) (m^2_{A'}+ 2m_e E_e-2m_e^2)^2}
\end{equation}
for the pseudo-Dirac elastic DM model,
\begin{equation}
    \frac{\dd \sigma (\chi e\to \chi e)}{\dd E_e}
    =
      4\pi \epsilon^2 \alpha \alpha_D
    \frac{
      2m_e E^2_{\chi}-(2m_e E_{\chi} +m_{\chi}^2)(E_e-m_e)
    }{(E_{\chi}^2-m_{\chi}^2) (m^2_{A'}+ 2m_e E_e-2m_e^2)^2}
\end{equation}
for the scalar elastic DM model, and
\begin{align}
\frac{\dd \sigma (\chi e\to \chi e)}{\dd E_e}=4\pi \epsilon^2 \alpha \alpha_D^{} \frac{2 m_e (E^2_{\chi} -m_{\chi}^2)+[m_{\chi}^2 -m_e (2 E_{\chi}-E_e +2 m_e)](E_e-m_e)}{(E^2_{\chi} -m_{\chi}^2)(m_{A'}^2 +2 m_e E_e -2m_e^2)^2}
\end{align}
for the Majorana DM model.
The range of $E_e$ is given by Eq.~\eqref{eq:minrecoil} with replacing $\chi_i$ by $\chi$.

\section{Neutrino-induced background }
\label{app:neucross}
We evaluate the number of neutrino-induced irreducible and reducible BG events from the neutrino flux of Fig.~\ref{fig:nu}.
The number of irreducible BG events does not depend on the details of the detector design and is evaluated in the same way as the signal rate calculation in Section~\ref{sec:setup}.
On the other hand, the number of reducible BG events highly depends on the detector design, so we conservatively take a pessimistic scenario without imposing the threshold energy of the electromagnetic showers\footnote{The rate of quasi-elastic scattering and pion production processes would not change significantly by imposing the threshold $E_{\rm min}=1$ GeV because their cross sections are very small for neutrino energy smaller than 1 GeV.}. 
This section describes the detailed calculations of the number of BG events.

\subsection{Irreducible background}

\begin{figure}[t]
\centering
\begin{subfigure}{0.43\textwidth}%
\includegraphics[width=\textwidth]{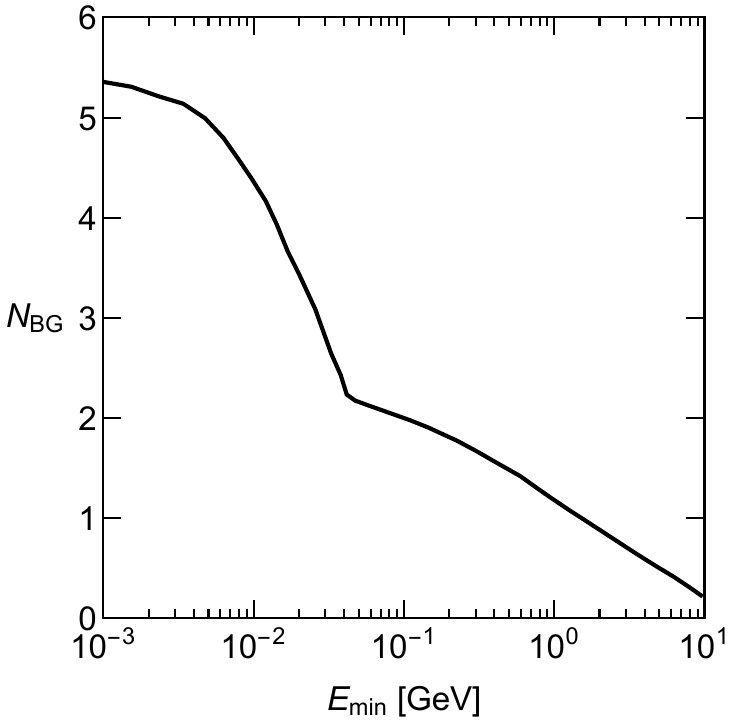}
\end{subfigure}%
\begin{subfigure}{0.45\textwidth}%
\includegraphics[width=\textwidth]{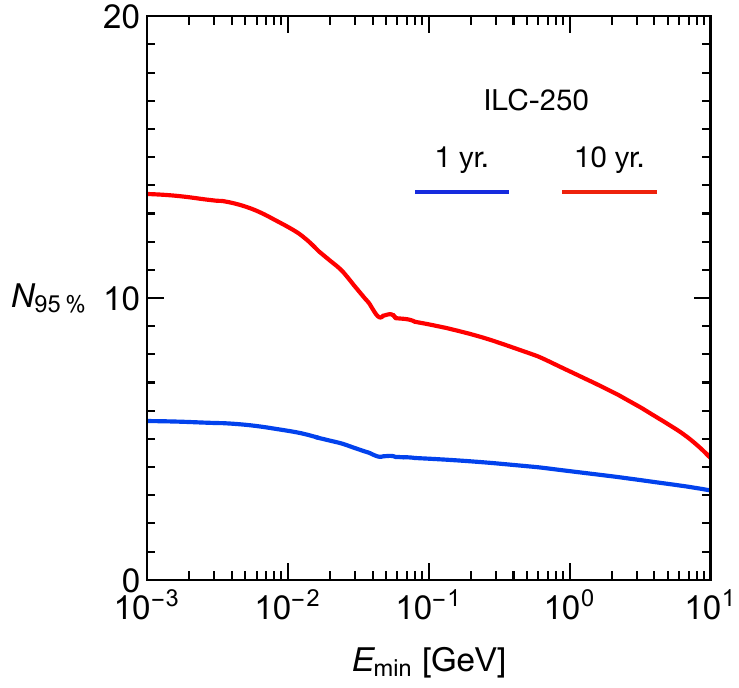}
\end{subfigure}%
\caption{
Left panel: The number of irreducible BG events in a 1-year run as a function of the threshold $E_{\rm min}$.
Right panel: The number of signal events at 95\% C.L. in a 1-year (blue) and 10-year (red) run as a function of the threshold $E_{\rm min}$.
}
\label{fig:bemre}
\end{figure}

The number of neutrino-electron recoil events is expressed by the following formula:
\begin{align}
    N_{\rm BG}=N_{e^{\pm}}\sum_{i=\nu_{e,\mu,\tau},\bar{\nu}_{e,\mu,\tau}}\int \dd E_{i}\, \pi r_{\rm det}^2\cdot\frac{\dd\phi_{i}}{\dd E_{i}}\cdot n^{\rm (det)}_{e^-}\cdot l_{\rm det}\cdot \int_{E_{{\rm min}}}\dd E_R \frac{\dd\sigma(i e^-\to i e^-)}{\dd E_R},
\end{align}
where $\phi_{i}$ is the neutrino flux at the detector position (Fig.~\ref{fig:nu} right) and $\sigma(i e^-\to i e^-)$ denotes the neutrino-electron elastic scattering cross section~\cite{Marciano:2003eq,Formaggio:2012cpf}.

In the left panel of Fig.~\ref{fig:bemre}, the number of neutrino-electron recoil events per year is shown as a function of the threshold $E_{\rm min}$.
The rate is reduced to $N^{\rm BG}_{\rm beam}\sim 1/$year by imposing $E_{\rm min}=1~{\rm GeV}$.
By taking into account the irreducible BG events and assuming no signal events are observed, we estimate the expected upper bounds on the number of signal events at 95\% C.L.\footnote{The Poisson 95\% C.L. upper bound corresponds to three signal events in the absence of BG events.}, shown by the solid lines in the right panel of Fig.~\ref{fig:bemre}.

\subsection{Reducible background}

The number of quasi-elastic  scattering events is estimated by the following formula:
\begin{align}
    N_{{\rm QE}}&=N_{e^{\pm}}\int \dd E_{\nu}\, \pi r_{\rm det}^2\cdot  l_{\rm det}\cdot\left( 
        \frac{\dd\phi_{\nu_{e}}}{\dd E_{\nu}}\,n^{\rm (det)}_{n}\, \sigma(\nu_e n \to e^- p)+\frac{\dd\phi_{\bar{\nu}_{e}}}{\dd E_{\nu}}\, n^{\rm (det)}_{p}\, \sigma(\bar{\nu}_e p \to e^+ n)\right),\label{eq:QEform}
\end{align}
where $\phi_{\nu_e}$ and $\phi_{\bar{\nu}_e}$ are the electron-neutrino and electron-antineutrino fluxes at the detector position (see the right panel of Fig.~\ref{fig:nu}), $n^{\rm (det)}_{n(p)}$ is the neutron (proton) number density of the detector, and $\sigma$ denotes the quasi-elastic scattering cross section~\cite{Formaggio:2012cpf}.
In contrast to the estimation of the irreducible BG events, we do not impose the threshold energy of the electromagnetic shower; Eq.~\eqref{eq:QEform} is therefore a conservative estimate.
From Eq.~\eqref{eq:QEform}, the number of the quasi-elastic scattering events per year is estimated as
\begin{align}
    N_{\rm QE}\sim 2\times 10^2 \times \biggl(\frac{N_{e^{\pm}}}{4\times 10^{21}}\biggr)\biggl(\frac{r_{\rm det}}{2~{\rm m}}\biggr)^2 \biggl(\frac{l_{\rm det}}{0.64~{\rm m}}\biggr).
\end{align}

Similarly, the number of the single-$\pi^0$ production events is given by
\begin{equation}
\begin{split}
   N_{{\rm pion}}
  &=
  N_{e^{\pm}}\int \dd E_{\nu}\, \pi r_{\rm det}^2\cdot  l_{\rm det}
  \\&\quad
    \times
  \Biggl\{
\frac{\dd\phi_{\nu_{\mu}}}{\dd E_{\nu}}
    \biggl[
      n^{\rm (det)}_{n} \Bigl(\sigma(\nu_{\mu} n \to \mu^- p \pi^0)+\sigma(\nu_{\mu} n \to \nu_{\mu} n \pi^0)\Bigr)
 +n^{\rm (det)}_{p} \sigma(\nu_{\mu} p \to \nu_{\mu} p \pi^0)
    \biggr]
  \\&\qquad
    +\frac{\dd\phi_{\nu_{\bar{\mu}}}}{\dd E_{\nu}}
     \biggl[n^{\rm (det)}_{p} \Bigl(\sigma(\bar{\nu}_{\mu} p \to \mu^+ n \pi^0)+\sigma(\bar{\nu}_{\mu} p \to \bar{\nu}_{\mu} p \pi^0)\Bigr)
 +n^{\rm (det)}_{n} \sigma(\bar{\nu}_{\mu} n \to \bar{\nu}_{\mu} n \pi^0)\biggr]
    \Biggr\},
\end{split}
\end{equation}
where the single-$\pi^0$ production cross sections are listed in Ref.~\cite{Formaggio:2012cpf}.
Combining the neutrino flux of the right panel of Fig.~\ref{fig:nu} and the cross sections in Ref.~\cite{Formaggio:2012cpf}, the number of the pion production events per year is obtained as
\begin{align}
    N_{\rm pion}\sim 2\times 10^3\times \biggl(\frac{N_{e^{\pm}}}{4\times 10^{21}}\biggr)\biggl(\frac{r_{\rm det}}{2~{\rm m}}\biggr)^2 \biggl(\frac{l_{\rm det}}{0.64~{\rm m}}\biggr).
\end{align}
As in the calculation of $N\w{QE}$, the threshold energy of the electromagnetic shower is not imposed and thus this estimate is conservative.

\bibliographystyle{utphys28mod}
{\small
\bibliography{ref}

\providecommand{\href}[2]{#2}\begingroup\begin{thebibliography}{10}

\bibitem{Baer:2013cma}
H.~Baer \emph{et al}., eds., ``{The International Linear Collider Technical
  Design Report - Volume 2: Physics}.'' {\ttfamily
  \href{https://arxiv.org/abs/1306.6352}{arXiv:1306.6352}}.

\bibitem{Fujii:2017vwa}
K.~Fujii \emph{et al}., ``{Physics Case for the 250 GeV Stage of the
  International Linear Collider}.'' {\ttfamily
  \href{https://arxiv.org/abs/1710.07621}{arXiv:1710.07621}}.

\bibitem{ILCInternationalDevelopmentTeam:2022izu}
{\bfseries ILC International Development Team} Collaboration, ``{The
  International Linear Collider: Report to Snowmass 2021}.'' {\ttfamily
  \href{https://arxiv.org/abs/2203.07622}{arXiv:2203.07622}}.

\bibitem{Kanemura:2015cxa}
S.~Kanemura, T.~Moroi, and T.~Tanabe, ``{Beam dump experiment at future
  electron\textendash{}positron colliders},''
  \href{https://dx.doi.org/10.1016/j.physletb.2015.10.002}{Phys.\  Lett.\  B
  {\bfseries 751} (2015) 25--28} {\ttfamily
  [\href{https://arxiv.org/abs/1507.02809}{arXiv:1507.02809}]}.

\bibitem{Sakaki:2020mqb}
Y.~Sakaki and D.~Ueda, ``{Searching for new light particles at the
  international linear collider main beam dump},''
  \href{https://dx.doi.org/10.1103/PhysRevD.103.035024}{Phys.\  Rev.\  D
  {\bfseries 103} (2021) 035024} {\ttfamily
  [\href{https://arxiv.org/abs/2009.13790}{arXiv:2009.13790}]}.

\bibitem{Asai:2021ehn}
K.~Asai, S.~Iwamoto, Y.~Sakaki, and D.~Ueda, ``{New physics searches at the ILC
  positron and electron beam dumps},''
  \href{https://dx.doi.org/10.1007/JHEP09(2021)183}{JHEP {\bfseries 09} (2021)
  183} {\ttfamily [\href{https://arxiv.org/abs/2105.13768}{arXiv:2105.13768}]}.

\bibitem{Asai:2021xtg}
K.~Asai, T.~Moroi, and A.~Niki, ``{Leptophilic Gauge Bosons at ILC Beam Dump
  Experiment},''
  \href{https://dx.doi.org/10.1016/j.physletb.2021.136374}{Phys.\  Lett.\  B
  {\bfseries 818} (2021) 136374} {\ttfamily
  [\href{https://arxiv.org/abs/2104.00888}{arXiv:2104.00888}]}.

\bibitem{Moroi:2022qwz}
T.~Moroi and A.~Niki, ``{Leptophilic Gauge Bosons at Lepton Beam Dump
  Experiments}.'' {\ttfamily
  \href{https://arxiv.org/abs/2205.11766}{arXiv:2205.11766}}.

\bibitem{Nojiri:2022xqn}
M.~M.~Nojiri, Y.~Sakaki, K.~Tobioka, and D.~Ueda, ``{First evaluation of meson
  and \ensuremath{\tau} lepton spectra and search for heavy neutral leptons at
  ILC beam dump},'' \href{https://dx.doi.org/10.1007/JHEP12(2022)145}{JHEP
  {\bfseries 12} (2022) 145} {\ttfamily
  [\href{https://arxiv.org/abs/2206.13523}{arXiv:2206.13523}]}.

\bibitem{Giffin:2022rei}
P.~Giffin, S.~Gori, Y.-D.~Tsai, and D.~Tuckler, ``{Heavy Neutral Leptons at
  Beam Dump Experiments of Future Lepton Colliders}.'' {\ttfamily
  \href{https://arxiv.org/abs/2206.13745}{arXiv:2206.13745}}.

\bibitem{PADME:2023vvr}
{\bfseries PADME} Collaboration, ``{Status and Prospects of PADME},'' in {\em
  {57th Rencontres de Moriond on Electroweak Interactions and Unified
  Theories}}.
\newblock 2023.
\newblock {\ttfamily
  \href{https://arxiv.org/abs/2305.08684}{arXiv:2305.08684}}.

\bibitem{Alexander:2016aln}
J.~Alexander \emph{et al}., ``{Dark Sectors 2016 Workshop: Community Report}.''
  {\ttfamily \href{https://arxiv.org/abs/1608.08632}{arXiv:1608.08632}}.

\bibitem{Battaglieri:2017aum}
M.~Battaglieri \emph{et al}., ``{US Cosmic Visions: New Ideas in Dark Matter
  2017: Community Report},'' in {\em {U.S. Cosmic Visions: New Ideas in Dark
  Matter}}.
\newblock 2017.
\newblock {\ttfamily
  \href{https://arxiv.org/abs/1707.04591}{arXiv:1707.04591}}.

\bibitem{Tucker-Smith:2001myb}
D.~Tucker-Smith and N.~Weiner, ``{Inelastic dark matter},''
  \href{https://dx.doi.org/10.1103/PhysRevD.64.043502}{Phys.\  Rev.\  D
  {\bfseries 64} (2001) 043502} {\ttfamily
  [\href{https://arxiv.org/abs/hep-ph/0101138}{hep-ph/0101138}]}.

\bibitem{BDX:2016akw}
{\bfseries BDX} Collaboration, ``{Dark Matter Search in a Beam-Dump eXperiment
  (BDX) at Jefferson Lab}.'' {\ttfamily
  \href{https://arxiv.org/abs/1607.01390}{arXiv:1607.01390}}.

\bibitem{Satyamurthy:2012zz}
P.~Satyamurthy, P.~Rai, V.~Tiwari, K.~Kulkarni, \emph{et al}., ``{Design of an
  18-MW vortex flow water beam dump for 500-GeV electrons/positrons of an
  international linear collider},''
  \href{https://dx.doi.org/10.1016/j.nima.2012.01.075}{Nucl.\  Instrum.\
  Meth.\  A {\bfseries 679} (2012) 67--81}.

\bibitem{Behnke:2013xla}
T.~Behnke \emph{et al}., eds., ``{The International Linear Collider Technical
  Design Report---Volume 1: Executive Summary}.'' {\ttfamily
  \href{https://arxiv.org/abs/1306.6327}{arXiv:1306.6327}}.

\bibitem{Adolphsen:2013jya}
C.~Adolphsen \emph{et al}., eds., ``{The International Linear Collider
  Technical Design Report---Volume~3.I: Accelerator R\&D in the Technical
  Design Phase}.'' {\ttfamily
  \href{https://arxiv.org/abs/1306.6353}{arXiv:1306.6353}}.

\bibitem{Adolphsen:2013kya}
C.~Adolphsen \emph{et al}., eds., ``{The International Linear Collider
  Technical Design Report---Volume~3.II: Accelerator Baseline Design}.''
  {\ttfamily \href{https://arxiv.org/abs/1306.6328}{arXiv:1306.6328}}.

\bibitem{Behnke:2013lya}
T.~Behnke \emph{et al}., eds., ``{The International Linear Collider Technical
  Design Report---Volume 4: Detectors}.'' {\ttfamily
  \href{https://arxiv.org/abs/1306.6329}{arXiv:1306.6329}}.

\bibitem{Asai:2022zxw}
K.~Asai, A.~Das, J.~Li, T.~Nomura, and O.~Seto, ``{Chiral Z' in FASER, FASER2,
  DUNE, and ILC beam dump experiments},''
  \href{https://dx.doi.org/10.1103/PhysRevD.106.095033}{Phys.\  Rev.\  D
  {\bfseries 106} (2022) 095033} {\ttfamily
  [\href{https://arxiv.org/abs/2206.12676}{arXiv:2206.12676}]}.

\bibitem{Hirayama:2005zm}
H.~Hirayama, Y.~Namito, A.~F.~Bielajew, S.~J.~Wilderman, and W.~R.~Nelson,
  ``{The EGS5 code system}.''.

\bibitem{Sato:2018}
T.~Sato, Y.~Iwamoto, S.~Hashimoto, T.~Ogawa, \emph{et al}., ``Features of
  particle and heavy ion transport code system (PHITS) version 3.02,'' Journal
  of Nuclear Science and Technology {\bfseries 55} (2018) 684--690.

\bibitem{Agostinelli:2002hh}
{\bfseries GEANT4} Collaboration, ``{GEANT4--a simulation toolkit},''
  \href{https://dx.doi.org/10.1016/S0168-9002(03)01368-8}{Nucl.\  Instrum.\
  Meth.\  A {\bfseries 506} (2003) 250--303}.

\bibitem{Bierlich:2022pfr}
C.~Bierlich \emph{et al}., ``{A comprehensive guide to the physics and usage of
  PYTHIA 8.3}.'' {\ttfamily
  \href{https://arxiv.org/abs/2203.11601}{arXiv:2203.11601}}.

\bibitem{SHiP:2015vad}
{\bfseries SHiP} Collaboration, ``{A facility to Search for Hidden Particles
  (SHiP) at the CERN SPS}.'' {\ttfamily
  \href{https://arxiv.org/abs/1504.04956}{arXiv:1504.04956}}.

\bibitem{Allison:2006ve}
J.~Allison \emph{et al}., ``{Geant4 developments and applications},''
  \href{https://dx.doi.org/10.1109/TNS.2006.869826}{IEEE Trans.\  Nucl.\  Sci.\
   {\bfseries 53} (2006) 270}.

\bibitem{sato2006analytical}
T.~Sato and K.~Niita, ``Analytical functions to predict cosmic-ray neutron
  spectra in the atmosphere,'' Radiation research {\bfseries 166} (2006)
  544--555.

\bibitem{sato2008development}
T.~Sato, H.~Yasuda, K.~Niita, A.~Endo, and L.~Sihver, ``Development of PARMA:
  PHITS-based analytical radiation model in the atmosphere,'' Radiation
  research {\bfseries 170} (2008) 244--259.

\bibitem{sato2015analytical}
T.~Sato, ``Analytical model for estimating terrestrial cosmic ray fluxes nearly
  anytime and anywhere in the world: Extension of PARMA/EXPACS,'' PloS one
  {\bfseries 10} (2015) e0144679.

\bibitem{Leane:2018kjk}
R.~K.~Leane, T.~R.~Slatyer, J.~F.~Beacom, and K.~C.~Y.~Ng, ``{GeV-scale thermal
  WIMPs: Not even slightly ruled out},''
  \href{https://dx.doi.org/10.1103/PhysRevD.98.023016}{Phys.\  Rev.\  D
  {\bfseries 98} (2018) 023016} {\ttfamily
  [\href{https://arxiv.org/abs/1805.10305}{arXiv:1805.10305}]}.

\bibitem{Filimonova:2022pkj}
A.~Filimonova, S.~Junius, L.~Lopez~Honorez, and S.~Westhoff, ``{Inelastic Dirac
  dark matter},'' \href{https://dx.doi.org/10.1007/JHEP06(2022)048}{JHEP
  {\bfseries 06} (2022) 048} {\ttfamily
  [\href{https://arxiv.org/abs/2201.08409}{arXiv:2201.08409}]}.

\bibitem{Izaguirre:2017bqb}
E.~Izaguirre, Y.~Kahn, G.~Krnjaic, and M.~Moschella, ``{Testing Light Dark
  Matter Coannihilation With Fixed-Target Experiments},''
  \href{https://dx.doi.org/10.1103/PhysRevD.96.055007}{Phys.\  Rev.\  D
  {\bfseries 96} (2017) 055007} {\ttfamily
  [\href{https://arxiv.org/abs/1703.06881}{arXiv:1703.06881}]}.

\bibitem{Giudice:2017zke}
G.~F.~Giudice, D.~Kim, J.-C.~Park, and S.~Shin, ``{Inelastic Boosted Dark
  Matter at Direct Detection Experiments},''
  \href{https://dx.doi.org/10.1016/j.physletb.2018.03.043}{Phys.\  Lett.\  B
  {\bfseries 780} (2018) 543--552} {\ttfamily
  [\href{https://arxiv.org/abs/1712.07126}{arXiv:1712.07126}]}.

\bibitem{Marsicano:2018glj}
L.~Marsicano, M.~Battaglieri, M.~Bond\'\i{}, C.~D.~R.~Carvajal, \emph{et al}.,
  ``{Novel Way to Search for Light Dark Matter in Lepton Beam-Dump
  Experiments},''
  \href{https://dx.doi.org/10.1103/PhysRevLett.121.041802}{Phys.\  Rev.\
  Lett.\  {\bfseries 121} (2018) 041802} {\ttfamily
  [\href{https://arxiv.org/abs/1807.05884}{arXiv:1807.05884}]}.

\bibitem{Akesson:2022vza}
T.~\r{A}kesson \emph{et al}., ``{Current Status and Future Prospects for the
  Light Dark Matter eXperiment},'' in {\em {Snowmass 2021}}.
\newblock 2022.
\newblock {\ttfamily
  \href{https://arxiv.org/abs/2203.08192}{arXiv:2203.08192}}.

\bibitem{Berlin:2018bsc}
A.~Berlin, N.~Blinov, G.~Krnjaic, P.~Schuster, and N.~Toro, ``{Dark Matter,
  Millicharges, Axion and Scalar Particles, Gauge Bosons, and Other New Physics
  with LDMX},'' \href{https://dx.doi.org/10.1103/PhysRevD.99.075001}{Phys.\
  Rev.\  D {\bfseries 99} (2019) 075001} {\ttfamily
  [\href{https://arxiv.org/abs/1807.01730}{arXiv:1807.01730}]}.

\bibitem{Izaguirre:2015zva}
E.~Izaguirre, G.~Krnjaic, and B.~Shuve, ``{Discovering Inelastic Thermal-Relic
  Dark Matter at Colliders},''
  \href{https://dx.doi.org/10.1103/PhysRevD.93.063523}{Phys.\  Rev.\  D
  {\bfseries 93} (2016) 063523} {\ttfamily
  [\href{https://arxiv.org/abs/1508.03050}{arXiv:1508.03050}]}.

\bibitem{Erler:2000jg}
T.~Behnke \emph{et al}., eds., ``{Physics impact of GigaZ},''
  \href{https://dx.doi.org/10.1016/S0370-2693(00)00749-8}{Phys.\  Lett.\  B
  {\bfseries 486} (2000) 125--133} {\ttfamily
  [\href{https://arxiv.org/abs/hep-ph/0005024}{hep-ph/0005024}]}.

\bibitem{Cui:2009xq}
Y.~Cui, D.~E.~Morrissey, D.~Poland, and L.~Randall, ``{Candidates for Inelastic
  Dark Matter},'' \href{https://dx.doi.org/10.1088/1126-6708/2009/05/076}{JHEP
  {\bfseries 05} (2009) 076} {\ttfamily
  [\href{https://arxiv.org/abs/0901.0557}{arXiv:0901.0557}]}.

\bibitem{Cheung:2009qd}
C.~Cheung, J.~T.~Ruderman, L.-T.~Wang, and I.~Yavin, ``{Kinetic Mixing as the
  Origin of Light Dark Scales},''
  \href{https://dx.doi.org/10.1103/PhysRevD.80.035008}{Phys.\  Rev.\  D
  {\bfseries 80} (2009) 035008} {\ttfamily
  [\href{https://arxiv.org/abs/0902.3246}{arXiv:0902.3246}]}.

\bibitem{Arina:2011cu}
C.~Arina and N.~Sahu, ``{Asymmetric Inelastic Inert Doublet Dark Matter from
  Triplet Scalar Leptogenesis},''
  \href{https://dx.doi.org/10.1016/j.nuclphysb.2011.09.014}{Nucl.\  Phys.\  B
  {\bfseries 854} (2012) 666--699} {\ttfamily
  [\href{https://arxiv.org/abs/1108.3967}{arXiv:1108.3967}]}.

\bibitem{Okada:2019sbb}
N.~Okada and O.~Seto, ``{Inelastic extra $U(1)$ charged scalar dark matter},''
  \href{https://dx.doi.org/10.1103/PhysRevD.101.023522}{Phys.\  Rev.\  D
  {\bfseries 101} (2020) 023522} {\ttfamily
  [\href{https://arxiv.org/abs/1908.09277}{arXiv:1908.09277}]}.

\bibitem{Batell:2014mga}
B.~Batell, R.~Essig, and Z.~Surujon, ``{Strong Constraints on Sub-GeV Dark
  Sectors from SLAC Beam Dump E137},''
  \href{https://dx.doi.org/10.1103/PhysRevLett.113.171802}{Phys.\  Rev.\
  Lett.\  {\bfseries 113} (2014) 171802} {\ttfamily
  [\href{https://arxiv.org/abs/1406.2698}{arXiv:1406.2698}]}.

\bibitem{Andreev:2021fzd}
Y.~M.~Andreev \emph{et al}., ``{Improved exclusion limit for light dark matter
  from e+e- annihilation in NA64},''
  \href{https://dx.doi.org/10.1103/PhysRevD.104.L091701}{Phys.\  Rev.\  D
  {\bfseries 104} (2021) L091701} {\ttfamily
  [\href{https://arxiv.org/abs/2108.04195}{arXiv:2108.04195}]}.

\bibitem{deNiverville:2011it}
P.~deNiverville, M.~Pospelov, and A.~Ritz, ``{Observing a light dark matter
  beam with neutrino experiments},''
  \href{https://dx.doi.org/10.1103/PhysRevD.84.075020}{Phys.\  Rev.\  D
  {\bfseries 84} (2011) 075020} {\ttfamily
  [\href{https://arxiv.org/abs/1107.4580}{arXiv:1107.4580}]}.

\bibitem{Berlin:2018pwiSea}
A.~Berlin, S.~Gori, P.~Schuster, and N.~Toro, ``{Dark Sectors at the Fermilab
  SeaQuest Experiment},''
  \href{https://dx.doi.org/10.1103/PhysRevD.98.035011}{Phys.\  Rev.\  D
  {\bfseries 98} (2018) 035011} {\ttfamily
  [\href{https://arxiv.org/abs/1804.00661}{arXiv:1804.00661}]}.

\bibitem{MiniBooNEDM:2018cxm}
{\bfseries MiniBooNE DM} Collaboration, ``{Dark Matter Search in Nucleon, Pion,
  and Electron Channels from a Proton Beam Dump with MiniBooNE},''
  \href{https://dx.doi.org/10.1103/PhysRevD.98.112004}{Phys.\  Rev.\  D
  {\bfseries 98} (2018) 112004} {\ttfamily
  [\href{https://arxiv.org/abs/1807.06137}{arXiv:1807.06137}]}.

\bibitem{COHERENT:2021pvd}
{\bfseries COHERENT} Collaboration, ``{First Probe of Sub-GeV Dark Matter
  beyond the Cosmological Expectation with the COHERENT CsI Detector at the
  SNS},'' \href{https://dx.doi.org/10.1103/PhysRevLett.130.051803}{Phys.\
  Rev.\  Lett.\  {\bfseries 130} (2023) 051803} {\ttfamily
  [\href{https://arxiv.org/abs/2110.11453}{arXiv:2110.11453}]}.

\bibitem{Akimov:2022oyb}
D.~Akimov \emph{et al}., ``{The COHERENT Experimental Program},'' in {\em
  {Snowmass 2021}}.
\newblock 2022.
\newblock {\ttfamily
  \href{https://arxiv.org/abs/2204.04575}{arXiv:2204.04575}}.

\bibitem{CCM:2021leg}
{\bfseries CCM} Collaboration, ``{First dark matter search results from
  Coherent CAPTAIN-Mills},''
  \href{https://dx.doi.org/10.1103/PhysRevD.106.012001}{Phys.\  Rev.\  D
  {\bfseries 106} (2022) 012001} {\ttfamily
  [\href{https://arxiv.org/abs/2105.14020}{arXiv:2105.14020}]}.

\bibitem{Tsai:2019buq}
Y.-D.~Tsai, P.~deNiverville, and M.~X.~Liu, ``{Dark Photon and Muon $g-2$
  Inspired Inelastic Dark Matter Models at the High-Energy Intensity
  Frontier},'' \href{https://dx.doi.org/10.1103/PhysRevLett.126.181801}{Phys.\
  Rev.\  Lett.\  {\bfseries 126} (2021) 181801} {\ttfamily
  [\href{https://arxiv.org/abs/1908.07525}{arXiv:1908.07525}]}.

\bibitem{Izaguirre:2013uxa}
E.~Izaguirre, G.~Krnjaic, P.~Schuster, and N.~Toro, ``{New Electron Beam-Dump
  Experiments to Search for MeV to few-GeV Dark Matter},''
  \href{https://dx.doi.org/10.1103/PhysRevD.88.114015}{Phys.\  Rev.\  D
  {\bfseries 88} (2013) 114015} {\ttfamily
  [\href{https://arxiv.org/abs/1307.6554}{arXiv:1307.6554}]}.

\bibitem{Essig:2013vha}
R.~Essig, J.~Mardon, M.~Papucci, T.~Volansky, and Y.-M.~Zhong, ``{Constraining
  Light Dark Matter with Low-Energy $e^+e^-$ Colliders},''
  \href{https://dx.doi.org/10.1007/JHEP11(2013)167}{JHEP {\bfseries 11} (2013)
  167} {\ttfamily [\href{https://arxiv.org/abs/1309.5084}{arXiv:1309.5084}]}.

\bibitem{XENON:2019zpr}
{\bfseries XENON} Collaboration, ``{Search for Light Dark Matter Interactions
  Enhanced by the Migdal Effect or Bremsstrahlung in XENON1T},''
  \href{https://dx.doi.org/10.1103/PhysRevLett.123.241803}{Phys.\  Rev.\
  Lett.\  {\bfseries 123} (2019) 241803} {\ttfamily
  [\href{https://arxiv.org/abs/1907.12771}{arXiv:1907.12771}]}.

\bibitem{XENON:2021qze}
{\bfseries XENON} Collaboration, ``{Emission of single and few electrons in
  XENON1T and limits on light dark matter},''
  \href{https://dx.doi.org/10.1103/PhysRevD.106.022001}{Phys.\  Rev.\  D
  {\bfseries 106} (2022) 022001} {\ttfamily
  [\href{https://arxiv.org/abs/2112.12116}{arXiv:2112.12116}]}.

\bibitem{SENSEI:2020dpa}
{\bfseries SENSEI} Collaboration, ``{SENSEI: Direct-Detection Results on
  sub-GeV Dark Matter from a New Skipper-CCD},''
  \href{https://dx.doi.org/10.1103/PhysRevLett.125.171802}{Phys.\  Rev.\
  Lett.\  {\bfseries 125} (2020) 171802} {\ttfamily
  [\href{https://arxiv.org/abs/2004.11378}{arXiv:2004.11378}]}.

\bibitem{SuperCDMS:2020aus}
{\bfseries SuperCDMS} Collaboration, ``{Light Dark Matter Search with a
  High-Resolution Athermal Phonon Detector Operated Above Ground},''
  \href{https://dx.doi.org/10.1103/PhysRevLett.127.061801}{Phys.\  Rev.\
  Lett.\  {\bfseries 127} (2021) 061801} {\ttfamily
  [\href{https://arxiv.org/abs/2007.14289}{arXiv:2007.14289}]}.

\bibitem{SuperCDMS:2023sql}
{\bfseries SuperCDMS} Collaboration, ``{Search for low-mass dark matter via
  bremsstrahlung radiation and the Migdal effect in SuperCDMS},''
  \href{https://dx.doi.org/10.1103/PhysRevD.107.112013}{Phys.\  Rev.\  D
  {\bfseries 107} (2023) 112013} {\ttfamily
  [\href{https://arxiv.org/abs/2302.09115}{arXiv:2302.09115}]}.

\bibitem{EDELWEISS:2019vjv}
{\bfseries EDELWEISS} Collaboration, ``{Searching for low-mass dark matter
  particles with a massive Ge bolometer operated above-ground},''
  \href{https://dx.doi.org/10.1103/PhysRevD.99.082003}{Phys.\  Rev.\  D
  {\bfseries 99} (2019) 082003} {\ttfamily
  [\href{https://arxiv.org/abs/1901.03588}{arXiv:1901.03588}]}.

\bibitem{EDELWEISS:2022ktt}
{\bfseries EDELWEISS} Collaboration, ``{Search for sub-GeV dark matter via the
  Migdal effect with an EDELWEISS germanium detector with NbSi transition-edge
  sensors},'' \href{https://dx.doi.org/10.1103/PhysRevD.106.062004}{Phys.\
  Rev.\  D {\bfseries 106} (2022) 062004} {\ttfamily
  [\href{https://arxiv.org/abs/2203.03993}{arXiv:2203.03993}]}.

\bibitem{Belle-II:2018jsg}
{\bfseries Belle-II} Collaboration, ``{The Belle II Physics Book},''
  \href{https://dx.doi.org/10.1093/ptep/ptz106}{PTEP {\bfseries 2019} (2019)
  123C01} {\ttfamily
  [\href{https://arxiv.org/abs/1808.10567}{arXiv:1808.10567}]}. [Erratum: PTEP
  2020, 029201 (2020)].

\bibitem{Duerr:2019dmv}
M.~Duerr, T.~Ferber, C.~Hearty, F.~Kahlhoefer, \emph{et al}., ``{Invisible and
  displaced dark matter signatures at Belle II},''
  \href{https://dx.doi.org/10.1007/JHEP02(2020)039}{JHEP {\bfseries 02} (2020)
  039} {\ttfamily [\href{https://arxiv.org/abs/1911.03176}{arXiv:1911.03176}]}.

\bibitem{Kahn:2018cqs}
Y.~Kahn, G.~Krnjaic, N.~Tran, and A.~Whitbeck, ``{M$^{3}$: a new muon missing
  momentum experiment to probe (g \ensuremath{-} 2)$_{\mu}$ and dark matter at
  Fermilab},'' \href{https://dx.doi.org/10.1007/JHEP09(2018)153}{JHEP
  {\bfseries 09} (2018) 153} {\ttfamily
  [\href{https://arxiv.org/abs/1804.03144}{arXiv:1804.03144}]}.

\bibitem{Gninenko:2020hbd}
S.~N.~Gninenko, N.~V.~Krasnikov, and V.~A.~Matveev, ``{Search for dark sector
  physics with NA64},''
  \href{https://dx.doi.org/10.1134/S1063779620050044}{Phys.\  Part.\  Nucl.\
  {\bfseries 51} (2020) 829--858} {\ttfamily
  [\href{https://arxiv.org/abs/2003.07257}{arXiv:2003.07257}]}.

\bibitem{Schuster:2021mlr}
P.~Schuster, N.~Toro, and K.~Zhou, ``{Probing invisible vector meson decays
  with the NA64 and LDMX experiments},''
  \href{https://dx.doi.org/10.1103/PhysRevD.105.035036}{Phys.\  Rev.\  D
  {\bfseries 105} (2022) 035036} {\ttfamily
  [\href{https://arxiv.org/abs/2112.02104}{arXiv:2112.02104}]}.

\bibitem{Toups:2022yxs}
M.~Toups \emph{et al}., ``{PIP2-BD: GeV Proton Beam Dump at Fermilab's PIP-II
  Linac},'' in {\em {Snowmass 2021}}.
\newblock 2022.
\newblock {\ttfamily
  \href{https://arxiv.org/abs/2203.08079}{arXiv:2203.08079}}.

\bibitem{DUNE:2020lwj}
{\bfseries DUNE} Collaboration, ``{Deep Underground Neutrino Experiment (DUNE),
  Far Detector Technical Design Report, Volume I Introduction to DUNE},''
  \href{https://dx.doi.org/10.1088/1748-0221/15/08/T08008}{JINST {\bfseries 15}
  (2020) T08008} {\ttfamily
  [\href{https://arxiv.org/abs/2002.02967}{arXiv:2002.02967}]}.

\bibitem{DeRomeri:2019kic}
V.~De~Romeri, K.~J.~Kelly, and P.~A.~N.~Machado, ``{DUNE-PRISM Sensitivity to
  Light Dark Matter},''
  \href{https://dx.doi.org/10.1103/PhysRevD.100.095010}{Phys.\  Rev.\  D
  {\bfseries 100} (2019) 095010} {\ttfamily
  [\href{https://arxiv.org/abs/1903.10505}{arXiv:1903.10505}]}.

\bibitem{SHiP:2020noy}
{\bfseries SHiP} Collaboration, ``{Sensitivity of the SHiP experiment to light
  dark matter},'' \href{https://dx.doi.org/10.1007/JHEP04(2021)199}{JHEP
  {\bfseries 04} (2021) 199} {\ttfamily
  [\href{https://arxiv.org/abs/2010.11057}{arXiv:2010.11057}]}.

\bibitem{Berlin:2018jbm}
A.~Berlin and F.~Kling, ``{Inelastic Dark Matter at the LHC Lifetime Frontier:
  ATLAS, CMS, LHCb, CODEX-b, FASER, and MATHUSLA},''
  \href{https://dx.doi.org/10.1103/PhysRevD.99.015021}{Phys.\  Rev.\  D
  {\bfseries 99} (2019) 015021} {\ttfamily
  [\href{https://arxiv.org/abs/1810.01879}{arXiv:1810.01879}]}.

\bibitem{Kling:2022ykt}
F.~Kling, J.-L.~Kuo, S.~Trojanowski, and Y.-D.~Tsai, ``{FLArE up dark sectors
  with EM form factors at the LHC forward physics facility},''
  \href{https://dx.doi.org/10.1016/j.nuclphysb.2023.116103}{Nucl.\  Phys.\  B
  {\bfseries 987} (2023) 116103} {\ttfamily
  [\href{https://arxiv.org/abs/2205.09137}{arXiv:2205.09137}]}.

\bibitem{Berlin:2018pwi}
A.~Berlin, S.~Gori, P.~Schuster, and N.~Toro, ``{Dark Sectors at the Fermilab
  SeaQuest Experiment},''
  \href{https://dx.doi.org/10.1103/PhysRevD.98.035011}{Phys.\  Rev.\  D
  {\bfseries 98} (2018) 035011} {\ttfamily
  [\href{https://arxiv.org/abs/1804.00661}{arXiv:1804.00661}]}.

\bibitem{Jordan:2018gcd}
J.~R.~Jordan, Y.~Kahn, G.~Krnjaic, M.~Moschella, and J.~Spitz, ``{Signatures of
  Pseudo-Dirac Dark Matter at High-Intensity Neutrino Experiments},''
  \href{https://dx.doi.org/10.1103/PhysRevD.98.075020}{Phys.\  Rev.\  D
  {\bfseries 98} (2018) 075020} {\ttfamily
  [\href{https://arxiv.org/abs/1806.05185}{arXiv:1806.05185}]}.

\bibitem{Batell:2021ooj}
B.~Batell, J.~Berger, L.~Darm\'e, and C.~Frugiuele, ``{Inelastic dark matter at
  the Fermilab Short Baseline Neutrino Program},''
  \href{https://dx.doi.org/10.1103/PhysRevD.104.075026}{Phys.\  Rev.\  D
  {\bfseries 104} (2021) 075026} {\ttfamily
  [\href{https://arxiv.org/abs/2106.04584}{arXiv:2106.04584}]}.

\bibitem{SuperCDMS:2022kse}
{\bfseries SuperCDMS} Collaboration, ``{A Strategy for Low-Mass Dark Matter
  Searches with Cryogenic Detectors in the SuperCDMS SNOLAB Facility},'' in
  {\em {Snowmass 2021}}.
\newblock 2022.
\newblock {\ttfamily
  \href{https://arxiv.org/abs/2203.08463}{arXiv:2203.08463}}.

\bibitem{Marsicano:2018krp}
L.~Marsicano, M.~Battaglieri, M.~Bondi', C.~D.~R.~Carvajal, \emph{et al}.,
  ``{Dark photon production through positron annihilation in beam-dump
  experiments},'' \href{https://dx.doi.org/10.1103/PhysRevD.98.015031}{Phys.\
  Rev.\  D {\bfseries 98} (2018) 015031} {\ttfamily
  [\href{https://arxiv.org/abs/1802.03794}{arXiv:1802.03794}]}.

\bibitem{vonWeizsacker:1934nji}
C.~F.~von Weizsacker, ``{Radiation emitted in collisions of very fast
  electrons},'' \href{https://dx.doi.org/10.1007/BF01333110}{Z.\  Phys.\
  {\bfseries 88} (1934) 612--625}.

\bibitem{Williams:1935dka}
E.~J.~Williams, ``{Correlation of certain collision problems with radiation
  theory},'' Matematisk-fysiske Meddelelser {\bfseries 13} (1935) 1--50.

\bibitem{Kim:1973he}
K.~J.~Kim and Y.-S.~Tsai, ``{Improved Weizs\"acker-Williams Method and Its
  Application to Lepton and $W$-Boson Pair Production},''
  \href{https://dx.doi.org/10.1103/PhysRevD.8.3109}{Phys.\  Rev.\  D {\bfseries
  8} (1973) 3109}.

\bibitem{Tsai:1986tx}
S.~C.~Loken, ed., ``{AXION BREMSSTRAHLUNG BY AN ELECTRON BEAM},''
  \href{https://dx.doi.org/10.1103/PhysRevD.34.1326}{Phys.\  Rev.\  D
  {\bfseries 34} (1986) 1326}.

\bibitem{Liu:2017htz}
Y.-S.~Liu and G.~A.~Miller, ``{Validity of the Weizs\"acker-Williams
  approximation and the analysis of beam dump experiments: Production of an
  axion, a dark photon, or a new axial-vector boson},''
  \href{https://dx.doi.org/10.1103/PhysRevD.96.016004}{Phys.\  Rev.\  D
  {\bfseries 96} (2017) 016004} {\ttfamily
  [\href{https://arxiv.org/abs/1705.01633}{arXiv:1705.01633}]}.

\bibitem{Bjorken:2009mm}
J.~D.~Bjorken, R.~Essig, P.~Schuster, and N.~Toro, ``{New Fixed-Target
  Experiments to Search for Dark Gauge Forces},''
  \href{https://dx.doi.org/10.1103/PhysRevD.80.075018}{Phys.\  Rev.\  D
  {\bfseries 80} (2009) 075018} {\ttfamily
  [\href{https://arxiv.org/abs/0906.0580}{arXiv:0906.0580}]}.

\bibitem{Kim:2016zjx}
D.~Kim, J.-C.~Park, and S.~Shin, ``{Dark Matter
  \textquotedblleft{}Collider\textquotedblright{} from Inelastic Boosted Dark
  Matter},'' \href{https://dx.doi.org/10.1103/PhysRevLett.119.161801}{Phys.\
  Rev.\  Lett.\  {\bfseries 119} (2017) 161801} {\ttfamily
  [\href{https://arxiv.org/abs/1612.06867}{arXiv:1612.06867}]}.

\bibitem{Batell:2021blf}
B.~Batell, J.~L.~Feng, and S.~Trojanowski, ``{Detecting Dark Matter with
  Far-Forward Emulsion and Liquid Argon Detectors at the LHC},''
  \href{https://dx.doi.org/10.1103/PhysRevD.103.075023}{Phys.\  Rev.\  D
  {\bfseries 103} (2021) 075023} {\ttfamily
  [\href{https://arxiv.org/abs/2101.10338}{arXiv:2101.10338}]}.

\bibitem{Marciano:2003eq}
W.~J.~Marciano and Z.~Parsa, ``{Neutrino electron scattering theory},''
  \href{https://dx.doi.org/10.1088/0954-3899/29/11/013}{J.\  Phys.\  G
  {\bfseries 29} (2003) 2629--2645} {\ttfamily
  [\href{https://arxiv.org/abs/hep-ph/0403168}{hep-ph/0403168}]}.

\bibitem{Formaggio:2012cpf}
J.~A.~Formaggio and G.~P.~Zeller, ``{From eV to EeV: Neutrino Cross Sections
  Across Energy Scales},''
  \href{https://dx.doi.org/10.1103/RevModPhys.84.1307}{Rev.\  Mod.\  Phys.\
  {\bfseries 84} (2012) 1307--1341} {\ttfamily
  [\href{https://arxiv.org/abs/1305.7513}{arXiv:1305.7513}]}.

\end{thebibliography}\endgroup
}

\end{document}